\UseRawInputEncoding
\documentclass[preprint,
 amsmath,amssymb,
 aps,
]{revtex4-2}

\usepackage{graphicx}
\usepackage{dcolumn}
\usepackage{bm}

\usepackage{subcaption}
\usepackage{xcolor}
\usepackage{caption}

\begin{document}

\preprint{This manuscript has been submitted to Physical Review E for review}

\title{High-dimensional covariance matrix estimators on simulated portfolios with complex structures.}

\author{Andr\'es Garc\'ia-Medina}
\email{andgarm.n@gmail.com}
 \affiliation{Faculty of Sciences, Autonomous University of Baja California, Ensenada, 22860, Mexico.}%

\date{\today}

\begin{abstract}

We study the allocation of synthetic portfolios under hierarchical nested, one-factor, and diagonal structures of the population covariance matrix in a high-dimensional scenario. The noise reduction approaches for the sample realizations are based on random matrices, free probability, deterministic equivalents, and their combination with a data science hierarchical method known as two-step covariance estimators. The financial performance metrics from the simulations are compared with empirical data from companies comprising the S\&P 500 index using a moving window and walk-forward analysis.
The portfolio allocation strategies analyzed include the minimum variance portfolio (both with and without short-selling constraints) and the hierarchical risk parity approach. Our proposed hierarchical nested covariance model shows signatures of complex system interactions. The empirical financial data reproduces stylized portfolio facts observed in the complex and one-factor covariance models. The two-step estimators proposed here improve several financial metrics under the analyzed investment strategies.
The results pave the way for new risk management and diversification approaches when the number of assets is of the same order as the number of transaction days in the investment portfolio.
\end{abstract}

\maketitle


\section{Introduction}


Financial markets exhibit stylized facts that have been extensively studied by economists.
Physicists, meanwhile, have explored the complexity of financial markets through the lens of complex systems, giving rise to the emerging field of econophysics \cite{mantegna1999introduction}.
One of the most intriguing and challenging features to explain is the complex interactions between these markets. A common approach to quantifying these dependencies is through the covariance matrix, which plays a central role in portfolio theory.
This paper proposes a model for the covariance matrix that aims to capture this complexity, and its impact on asset allocation within an investment portfolio is analyzed.


Nevertheless, complex systems lack a formal mathematical definition but are categorized based on a series of characteristics.
The most distinctive of these is that when their constituent elements—in this case, financial markets—interact, new properties emerge that cannot be explained by the sum of their parts. In other words, their collective properties are not measurable within individual markets. Another characteristic of complex systems is the presence of hierarchical and heterogeneous structures among their constituents \cite{simon1962complexity, anderson1972more, barabasi2007architecture}.


A characteristic feature of these systems is that observable measures can often be represented by power laws. Indeed, one of the most active areas of research in econophysics is the study of power laws in financial markets under various configurations \cite{chakraborti2011econophysics, pereira2017econophysics}.
In particular, the authors of \cite{sanchez2021criticality} studied the eigenvalue spectrum of a two-dimensional Ising model and found that the scree plot follows a power law. This result has inspired the present work to propose a fully nested population model for the covariance matrix, which similarly exhibits a scree plot that approximately follows a power law.

This model thus possesses two distinctive features of complex systems, making it a suitable proxy for characterizing the complex interactions of financial markets. To our knowledge, this is the first time such a complex covariance matrix model has been proposed and applied. We conjecture that complex interactions, in a statistical sense, manifest in this covariance matrix structure at the population level.


On the other hand, the input data used to build the covariance matrix consists of market returns, which are assumed to be noisy realizations of a stochastic process. This noise is exacerbated by the temporal finiteness of the time series. Furthermore, when the number of assets $p$ is of the same order as the number of transaction days $n$, the classical estimator of the covariance matrix fails to accurately capture the population model. This scenario is known as the high-dimensionality limit and has been extensively studied by mathematicians, statisticians, and physicists. 

In practice, high dimensionality arises when we aim to diversify an investment portfolio by increasing $p$ while keeping $n$ relatively short to ensure a stationary set of time series. In this work, we implement covariance matrix estimators derived from high-dimensional statistics, leveraging techniques rooted in random matrix theory~(RMT) and data science~\cite{potters2020first, couillet2022random}.



In particular, we have considered rotational invariant estimators~(RIE) proposed and discussed in~\cite{ledoit2004well, ledoit2011eigenvectors, ledoit2020analytical, bun2017cleaning, ledoit2022power}, which are optimal with respect to the Frobenius loss function. We also implement variations of non-linear shrinkage estimators from the RIE family, which are optimal with respect to the Stein and Symmetrized Stein loss functions~\cite{ledoit2018optimal, ledoit2021shrinkage}. The non-linear shrinkage estimators have also been studied from the perspective of free probability in~\cite{speicher2012free, potters2020first}.  

Furthermore, we consider the estimator proposed in~\cite{yang2015robust}, which is built using modern RMT techniques via deterministic equivalents~\cite{speicher2012free, couillet2022random}. This estimator is optimized with respect to the realized variance of a minimum variance portfolio, serving as a loss function. Additionally, we include the two-step estimator method proposed in~\cite{garcia2023two} and implemented in a portfolio context in~\cite{garcia2024random}. In this way, two-step estimators were constructed for the non-linear shrinkage estimator variants and the estimator based on deterministic equivalents.


Thus, we are interested in studying the performance of high-dimensional covariance matrix estimators in the context of investment strategies. We analyze the weights of capital allocation for Markowitz minimum variance portfolios, both with and without short-selling restrictions~\cite{markowitz1952, roncalli2013introduction}, as well as for the hierarchical risk parity~(HRP) strategy proposed by De Prado~\cite{lopez2016building, de2020machine}.


The latter has been developed based on hierarchical clustering to bring the covariance matrix to a quasi-diagonal form. Then, the capital is allocated according to dendrogram linkages. This strategy seeks to diversify investment risk by performing allocation based on the cluster elements via bisection methods and Markowitz's theory.  
Hence, we applied the investment strategies and the estimators to realizations of three covariance structures: the completely nested covariance structure, a factor covariance model, and a diagonal covariance matrix. 
The first population covariance model is our proposed proxy to capture complex interactions. The second model is a factor model tested in~\cite{yang2015robust}, which, in terms of asset pricing models, represents the Sharpe Single Index Model~\cite{sharpe1964capital}. 
The Sharpe model has been shown to be a robust factor model for characterizing pricing formation from an RMT perspective~\cite{molero2023market}. 
The diagonal covariance matrix was originally proposed in~\cite{bai1998no} as a challenging population model to estimate from sample realizations.


Finally, we carry out the exercise of applying the estimators and investment strategies to empirical data. We consider the adjusted daily closing prices of the companies that make up the S\&P 500 index from 2012 to 2022 and calculate their returns.  

Using moving windows and walk-forward analysis, we analyze financial metrics such as the Herfindahl–Hirschman Index~($\mathcal{HHI}$), leverage~($\mathcal{L}$), Risk Diversification Index~($\mathcal{RDI}$), realized risk~($\mathcal{R}_{out}$), among others~\cite{tasche2008capital, roncalli2013introduction}, and compare their performance with respect to the simulations under controlled conditions.


We theoretically know that covariance estimators are optimal for specific loss functions under a given set of assumptions. However, we know little about their behavior outside these restrictions. This motivates us to perform a numerical comparison under challenging models of the covariance matrix and various investment strategies.  

In particular, we are interested in testing the proposed fully hierarchical model and determining the extent to which it serves as an appropriate proxy for capturing the complex structure of financial markets. The goal is to realistically characterize these dependencies through numerical comparisons.  

Our research question seeks to answer whether the performance of investment portfolios can reveal traces of the complex structures inherent in financial markets. Specifically, can the measurement noise due to high dimensionality be effectively eliminated by our proposed estimators to uncover the population structure of the data? Does the de-noising procedure help improve the performance of financial metrics?


We have found similarities in the performance of the empirical data in relation to the simulations for the fully nested covariance matrix models and the one-factor model. Thus, this study provides evidence to suggest that, from the perspective of portfolio theory, financial markets exhibit stylized facts that oscillate between a one-factor population model and a fully hierarchical dependency model.  

Further, the two-step estimator based on the deterministic equivalent estimator significantly reduces the financial metrics of capital diversification and leverage in both the synthetic models and the empirical data. This two-step estimator variant also improves turnover and maximum drawdown during the walk-forward analysis of the empirical data.  

For other metrics, variants of non-linear estimators and two-step estimators may also be suitable. Additionally, it is observed that among the investment strategies, HRP is the least sensitive to high-dimensional noise, meaning that the estimators have minimal impact on its performance.


In the following Section~2, the investment strategies are described. In Section~3, the different estimators of the covariance matrix implemented in this study are presented concisely. Section~4 outlines the characteristics of the proposed population models of the covariance matrix. In Section~5, the performance metrics used to evaluate the different combinations of estimators and strategies are described, both in the simulations and in the empirical data.  
Section~6 presents the methodology and results of the simulations, while Section~7 focuses on the empirical data. Finally, Section~8 discusses the main findings, and in Section~9, conclusions and proposals for future work are provided.

\section{Asset allocation models}
\label{allocation_models}

\subsection{Portfolio Theory}
Consider $p$ assets on $n$ trading days and denote by $s_{i,t}$ the price of asset $i=1,\dots,p$ at time $t=1,\dots,n$. The return $r_{i,t}$ is defined as
 \begin{equation}
 \label{returns}
 r_{i,t} = \frac{s_{i,t}-s_{i,t-1}}{s_{i,t-1}}
 \end{equation}
The amount of money invested in the asset $i$ is known as the portfolio weight and is given by the vector
\begin{equation}
  \mathbf{w} = (w_1,\dots,w_p)^T,
 \end{equation}
A positive weight represents a \emph{long position}, whereas a negative weight indicates a \emph{short position}.
Further, the expected return of the portfolio denoted here as $\mathcal{M}$ is defined as
\begin{equation}
\mathcal{M} = \mathbb{E}(\mathbf{w}^T \mathbf{r}) = \mathbf{w}^T\mathbb{E}(\mathbf{r}) = \mathbf{w}^T\mathbf{\mu},
\label{expected_profit}
\end{equation}
and the portfolio risk is expressed as a function of the population covariance matrix $\mathbf{\Sigma}$ of the returns
\begin{equation}
    \mathcal{R}^2 = \mathbf{w}^T\mathbf{\Sigma w}.
\end{equation}

\subsection{Minimum Variance Portfolio(MVP)}
The mean-variance allocation strategy of Markowitz\cite{markowitz1952} proposes to solve the following quadratic optimization problem to minimize the portfolio risk at a given level of expected return\cite{roncalli2013introduction}
\begin{equation} 
\underset{\mathbf{w}(\phi)\in\mathbb{R}^p}{max} \mathbf{w}^T \mathbf{\mu} - \frac{\phi}{2} \mathbf{w}^T \mathbf{\Sigma w}\quad\text{subject to}\quad \mathbf{1}^T\mathbf{w} = 1
\label{mean_variance}
\end{equation}
where $\phi$ is interpreted as the risk-aversion parameter.
Hence, if $\phi=\infty$ the problem is transformed to
\begin{equation} 
\underset{\mathbf{w}(\infty)\in\mathbb{R}^p}{min} \frac{1}{2} \mathbf{w}^T\mathbf{\Sigma w}\quad\text{subject to}\quad \mathbf{1}^T\mathbf{w} = 1
\label{mv}
\end{equation}
The reduced problem minimizes the volatility and is known as the Minimum Variance Portfolio~(MVP). Its solution is given by
\begin{equation}
  \hat{\mathbf{w}} = \frac{\mathbf{\Sigma}^{-1}\mathbf{1}}{\mathbf{1}^T\mathbf{\Sigma}^{-1} \mathbf{1}}.
  \label{optimal}
\end{equation}
Thus, the minimum risk associated with this optimal solution is given by
\begin{equation}
   \mathcal{R}^2_{true} = \frac{1}{\mathbf{1}^T \mathbf{\Sigma}^{-1}\mathbf{1}}
   \label{Risk_true}
  \end{equation}
However,  the population covariance matrix $\mathbf{\Sigma}$ is unknown \emph{a priori}. Consequently, the optimal solution is not achievable in any real situation.
Nevertheless,  it is possible to compute the in-sample risk $\mathcal{R}^2_{in}$ and the out-sample risk $\mathcal{R}^2_{out}$ using the expressions 
\begin{equation}
\begin{split}
& \mathcal{R}^2_{in} =  \frac{1}{\mathbf{1}^T \mathbf{S}_{in}^{-1}\mathbf{1}},\\
& \mathcal{R}^2_{out} =   \frac{\mathbf{1}^T\mathbf{S}^{-1}_{in}\mathbf{S}_{out}\mathbf{S}^{-1}_{in}\mathbf{1}}{(\mathbf{1}^T\mathbf{S}^{-1}_{in}\mathbf{1})^2},
\label{Risk_sample}
\end{split}
\end{equation}
where $\mathbf{S}_{in}$ and $\mathbf{S}_{out}$ are the in-sample and out-sample (realized) covariance matrices, respectively.


Further, it is possible to include the no short-selling restriction by considering the standard quadratic programming~(QP) problem. In this case, $\mathbf{w} \geq 0$, and a numerical solution can only be obtained~\cite{boyd2004convex}. This portfolio is also known as the long-only MVP because it does not allow short positions or negative weights. We will denote this portfolio as MVP+ in the subsequent analysis.

\subsection{Hierarchical Risk Parity (HRP)}
The hierarchical Risk Parity (HRP) is an asset allocation strategy proposed by De Prado~\cite{lopez2016building} that does not require the covariance matrix's inversion. The algorithm consists of three crucial steps. 

In the first step $\mathbf{\Sigma}$ is converted into a correlation matrix by  $\mathbf{C}= \mathbf{H}^{-1/2} \mathbf{S} \mathbf{H}^{-1/2}$, where $\mathbf{H}$ is a diagonal matrix of individual variances. 

Then, it is defined a distance measure $D_{i,j} = \sqrt{\frac{1}{2}(1-C_{i,j})}$, for $i,j=1,\dots,p$, where $p$ is the dimension of $\mathbf{\Sigma}$. 
Further, a distance of distances $\tilde{\mathbf{D}}$ is defined between the column vectors of $\mathbf{D}$
\begin{equation}
\tilde{D}_{i,j} = \sqrt{\sum_{k=1}^p(D_{k,i}-D_{k,j})^2},\quad i,j=1,\dots,p
\end{equation}
The first distance $D_{i,j}$ measures the dissimilarity between the assets $i,j$, while $\tilde{D}_{i,j}$ capture the closeness of the assets $i,j$ with the rest of the portfolio\cite{hudsonthames_hrp}. Hence, a dendrogram is built using the Single Linkage Clustering Analysis~(SLCA)\cite{johnson2002applied}.

The second step consists of applying a seriation algorithm to reorder the matrix rows and columns of $\mathbf{\Sigma}$ based on the dendrogram. The algorithm recursively replaces the clusters with their constituents until no cluster remains. The replacement preserves the order of the clustering and has the advantage of putting similar elements together, while dissimilar elements are placed apart. The output is a covariance matrix where stocks with larger covariance are arranged along the main diagonal and stocks with smaller variance are placed in the off-diagonal matrix. Thus, a quasi-diagonal covariance matrix is obtained.

The third and most important step is asset allocation using a recursive bisection algorithm. The idea is to define the variance of each cluster as the variance of an inverse-variance allocation and to split the allocation between adjacent clusters inversely proportional to the sum of their variance. The procedure makes use of the fact that  the standard quadratic optimization problem
of eq.~(\ref{mv}) has the particular solution
\begin{equation}
  \hat{\mathbf{w}} = \frac{diag(\mathbf{\Sigma})^{-1}}{tr(diag(\mathbf{\Sigma})^{-1})}.
  \label{optimal_particular}
\end{equation}
when $\mathbf{\Sigma}$ is diagonal. Then, given that the second step generates a quasi-diagonal covariance matrix De Prado proposes an algorithm based on (\ref{optimal_particular})to allocate capital\cite{lopez2016building}. The complete procedure is known as the HRP allocation strategy.

\section{Covariance estimators}
\label{estimators}

A \emph{naive estimator} of the population covariance matrix $\mathbf{\Sigma}$ given the empirical or sample covariance matrix $\mathbf{S}$ is given by
\begin{equation}
    \mathbf{\Xi}^{naive} = \mathbf{S},
\end{equation} 
which is a unbiased estimator of $\mathbf{\Sigma}$ whenever the number of variables $p$ is fixed and the number of observations $n\rightarrow\infty$\cite{johnson2002applied}.

An estimator that has roots in RMT but has been developed within the area of mathematical statistics with a Bayesian approach is the one proposed by Ledoit and Wolf\cite{ledoit2004well}.
They deal with the ill-conditioned problem of estimating a covariance matrix when $p$ grows at the same rate as $n$. The estimator is a convex linear combination of the sample covariance matrix with the identity matrix, which is known as \emph{linear shrinkage}
and is expressed as
\begin{equation}
            \mathbf{\Xi}^{linear} = \hat{\alpha} \zeta \mathbf{I} + (1-\hat{\alpha})\mathbf{S},
\end{equation}
where $\zeta=tr(\mathbf{S})/p$, and $\hat{\alpha}$ is the optimal parameter that shrinks $\mathbf{S}$ to the diagonal matrix $\zeta\mathbf{I}$. In this way, the estimator is the weighted average of the empirical covariance matrix and the matrix where all the variances are the same, whereas the covariances are equal to zero. The optimal value of $\hat{\alpha}$ is found to be asymptotically approximated by
  $\hat{\alpha}$~with respect to a quadratic loss function
\begin{equation}
     \hat{\alpha} = \frac{min\{\frac{1}{n} \sum_{i=1}^n|| \mathbf{X}_i \mathbf{X}_i^T - \mathbf{S} ||_{F}^2, ||\mathbf{S} - \mathbf{I}||_{F}^2\}}{||\mathbf{S} - \mathbf{I}||_{F}^2},
     \label{OptimalLS}
\end{equation}
where $|| \cdot||_{F}$ represents the Frobenius norm and $\mathbf{X}_i$ is the i-th column of the data matrix $\mathbf{X}$.  
  
A non-linear shrinkage formula minimizing the Frobenious loss functions has been proposed by\cite{ledoit2011eigenvectors} and studied from the point of view of free probability\cite{potters2020first}. The covariance matrix in high dimensions is estimated as

\begin{eqnarray}
\mathbf{\Xi}^{LP} &=& \sum_{k=1}^{p} \xi_k^{LP} v_k v_k^T,\quad
    \text{where}\\
\xi^{LP}_k  &=&  \lim_{\epsilon\rightarrow 0^{+}}\frac{\lambda_k}{|1-q+q\lambda_k G_S(\lambda_k-i\epsilon)|^2},
\label{LedoitPeche2011}    
\end{eqnarray}
where $\lambda_k, v_k$  are the eigenvalue and eigenvector tuple of $\mathbf{S}$, and $G_S$ is the Stieltjes transform of $\mathbf{S}$.

A variant of a non-linear shrinkage formula minimizing the Stein loss function has been derived in~\cite{ledoit2021shrinkage}. Here the estimated covariance function takes the form
\begin{eqnarray}
    \mathbf{\Xi}^{Stein} &=& \sum_{k=1}^{p} \xi_k^{Stein} v_k v_k^T,\quad
    \text{where}\\
    \xi^{Stein}_k  &=&  \lim_{\epsilon\rightarrow 0^{+}}\frac{\lambda_k}{1-q+2q\lambda_k Re[G_S(\lambda_k-i\epsilon)]}
\end{eqnarray}

Likewise, the non-linear shrinkage formula minimizing the Symmetrized Stein loss function leads to the estimated covariance matrix~\cite{ledoit2021shrinkage}

\begin{eqnarray}
    \mathbf{\Xi}^{SymStein} &=& \sum_{k=1}^{p} \xi_k^{SymStein} v_k v_k^T,\quad
    \text{where}\\
    \xi^{SymStein}_k  &=&  \sqrt{\xi^{LP}_k\xi^{Stein}_k}
\end{eqnarray}

In\cite{yang2015robust} the authors propose an online estimator based on the deterministic equivalent of the realized risk of the MVP.
A deterministic equivalent is an RMT technique of proof based on finite deterministic matrices having in probability (almost surely) the same scalar observations as the random ones\cite{couillet2022random}. In particular, $\mathbf{\Sigma}$ is estimated via the unique solution of the fixed-point equation
\begin{equation}
\mathbf{\Xi}^{YCM}(\hat{\rho}) = (1-\rho) \frac{1}{n}\sum_{i=1}^n \frac{\tilde{\mathbf{X}}_i\tilde{\mathbf{X}}_i^T}{\frac{1}{p}\tilde{\mathbf{X}}_i^T \mathbf{\Xi}^{YCM}(\hat{\rho})\tilde{\mathbf{X}}_i} + \rho \mathbf{I}_p
\end{equation}
where $\tilde{\mathbf{X}}_i = \mathbf{X}_i  - \frac{1}{n}\sum_{j=1}^n \mathbf{X}_j$, and the optimal $\hat{\rho}$ is obtained by
\begin{equation}
\hat{\rho} = \arg \min_{\rho \in [\epsilon+\max\{0,1-n/p\},1]}  \{ \hat{\sigma}^2_{sc}(\rho) \} 
\label{optimal_rho}
\end{equation}
where $\hat{\sigma}^2_{sc}(\rho)$ is a consistent estimation of the scaled realized portfolio risk via the deterministic equivalent of  $\mathcal{R}^2_{out}$\footnote{see \cite{yang2015robust} for the explicit mathematical expression}.

A novel approach to estimate the covariance matrix is proposed in \cite{tumminello2007kullback} using a hierarchical clustering algorithm. 
The procedures consist of transforming the empirical covariance matrix $\mathbf{S}$ into a correlation matrix $\mathbf{C}$.
Then, it is applied the transformation $\mathbf{D}=\mathbf{11}^T-\mathbf{C}$, which also satisfies the axioms of a distance measure, being  $\mathbf{1}$ a vector of ones of dimension $p$.
Then, it is constructed a dendrogram of $\mathbf{D}$ via the Average Linkage Clustering Analysis~(ALCA)\cite{johnson2002applied} and computed the distance $\rho$ between clusters at each hierarchical level. Hence, we can obtain a dissimilarity matrix  $\mathbf{D}(\rho)$ as a function of $\rho$, and the filtered covariance matrix is retrieved by the inverse transformation $\mathbf{\Xi}^{ALCA}=\mathbf{H}^{1/2}(\mathbf{11}^T-\mathbf{D}(\rho))\mathbf{H}^{1/2}$.

A state-of-the-art estimator is proposed in~\cite{garcia2023two} to deal with both the heterogeneous structure of the financial markets and the high-dimensional scenario. The core idea consists of applying as a first step a hierarchical estimator and then a variant of the RMT family. Here, we consider the combinations:
\begin{eqnarray}
    \mathbf{\Xi}^{2S(LP)} &:=& \mathbf{\Xi}^{LP}(\mathbf{\Xi}^{ALCA}).\\
    \mathbf{\Xi}^{2S(Stein)} &:=& \mathbf{\Xi}^{Stein}(\mathbf{\Xi}^{ALCA}).\\
    \mathbf{\Xi}^{2S(SymStein)} &:=& \mathbf{\Xi}^{SymStein}(\mathbf{\Xi}^{ALCA}).\\
    \mathbf{\Xi}^{2S(YCM)} &:=& \mathbf{\Xi}^{YCM}(\mathbf{\Xi}^{ALCA}).
\end{eqnarray}

\section{Covariance models}
\label{covariance_models}
We consider a multiplicative noise model with the following structure
\begin{eqnarray}
    \mathbf{Y} = \sqrt{\mathbf{\Sigma}}\mathbf{X}\\
    \mathbf{S} = \frac{1}{n} \sqrt{\mathbf{\Sigma}} \mathbf{X}\mathbf{X}^T\sqrt{\mathbf{\Sigma}}
    \label{noise_model}
\end{eqnarray}
where $\mathbf{Y}$ is the $p\times n$ data matrix, $\mathbf{\Sigma}$ is the $p\times p$ population covariance matrix.

\begin{figure}[hbtp]
    \centering
    \begin{subfigure}[b]{0.45\textwidth}
       \includegraphics[scale=0.25]{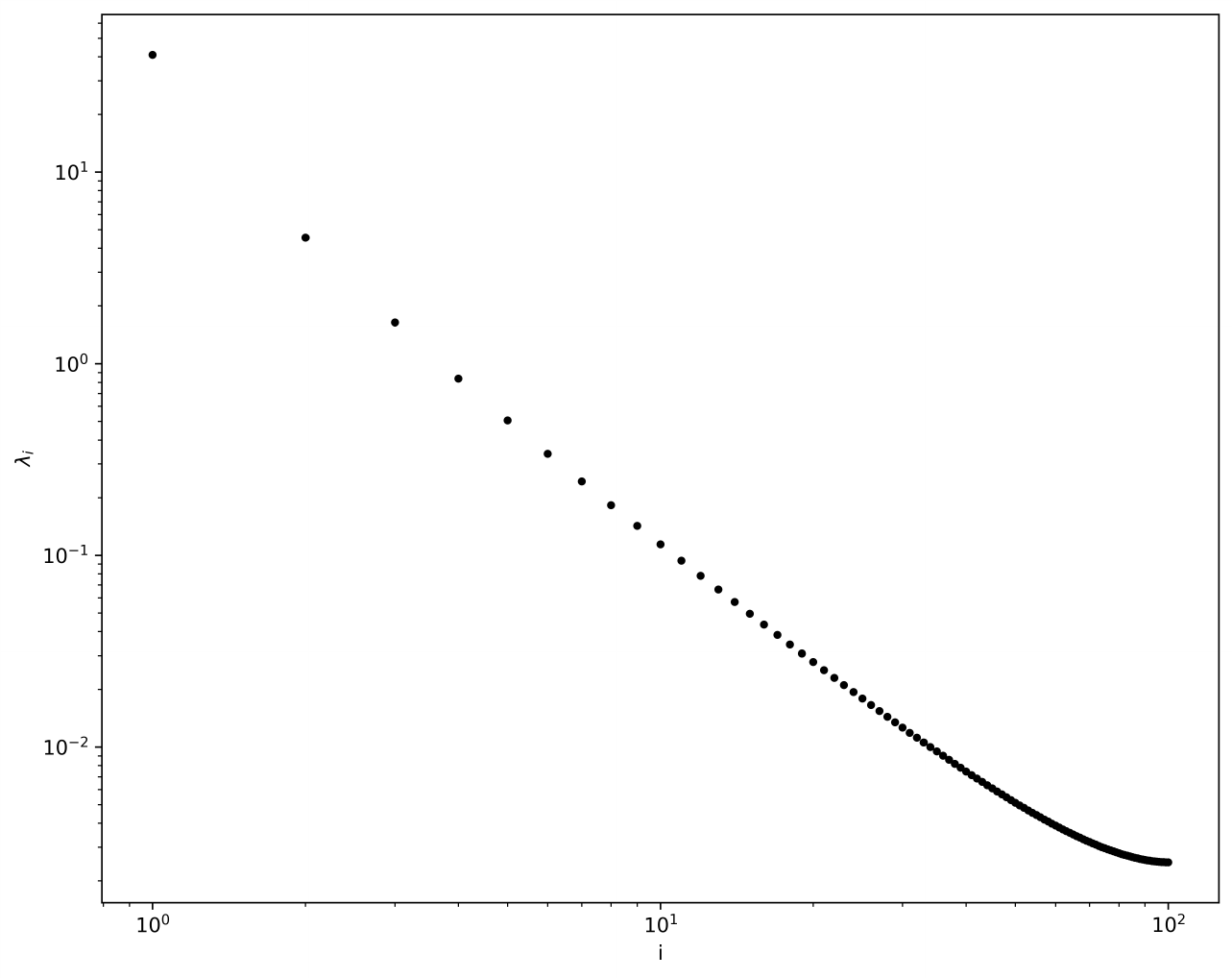}
        \caption{}
    \end{subfigure}
    \begin{subfigure}[b]{0.45\textwidth}
        \includegraphics[scale=0.25]{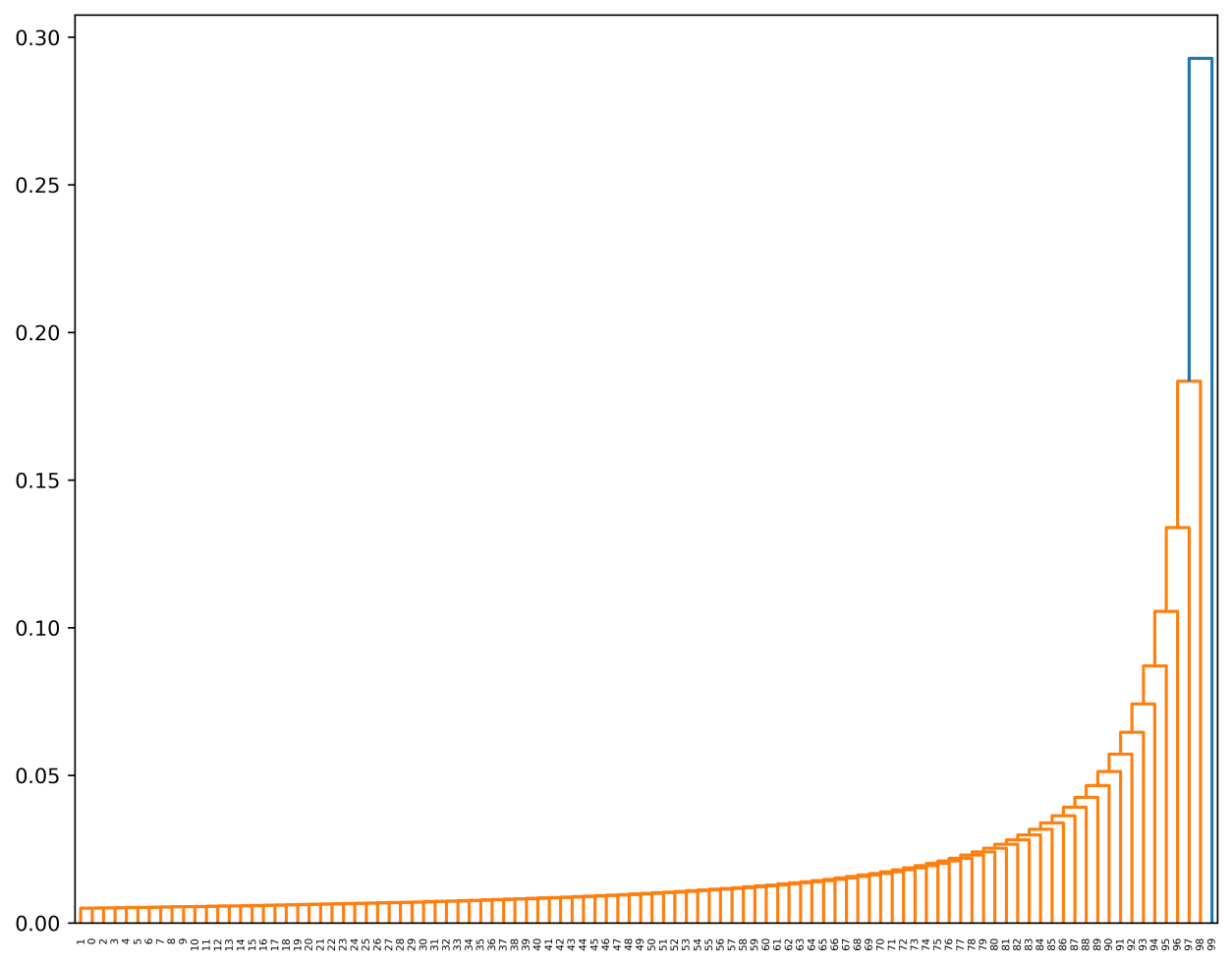}
        \caption{}
    \end{subfigure}\\
    \begin{subfigure}[b]{0.45\textwidth}
        \includegraphics[scale=0.25]{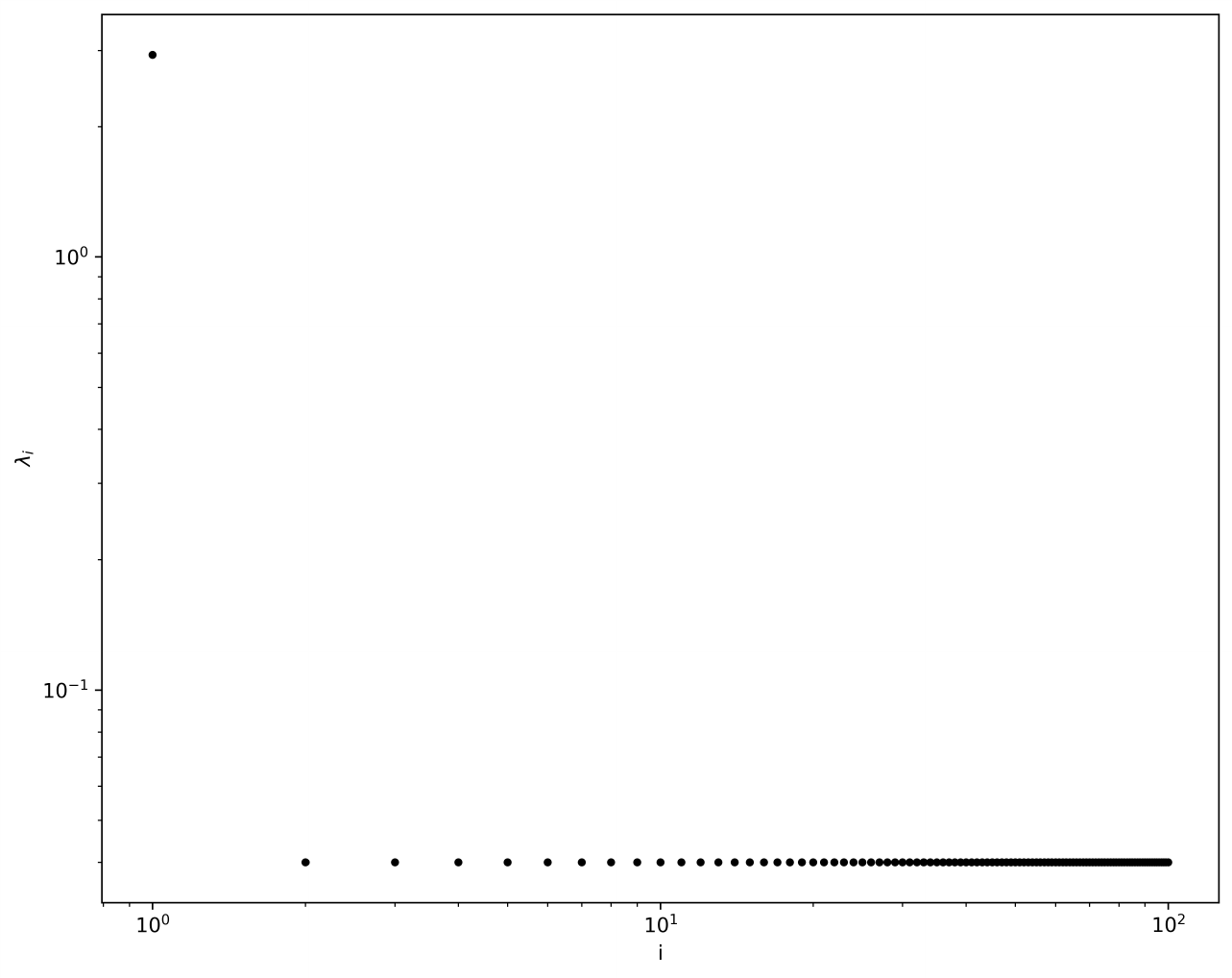}
        \caption{}
    \end{subfigure}
    \begin{subfigure}[b]{0.45\textwidth}
        \includegraphics[scale=0.25]{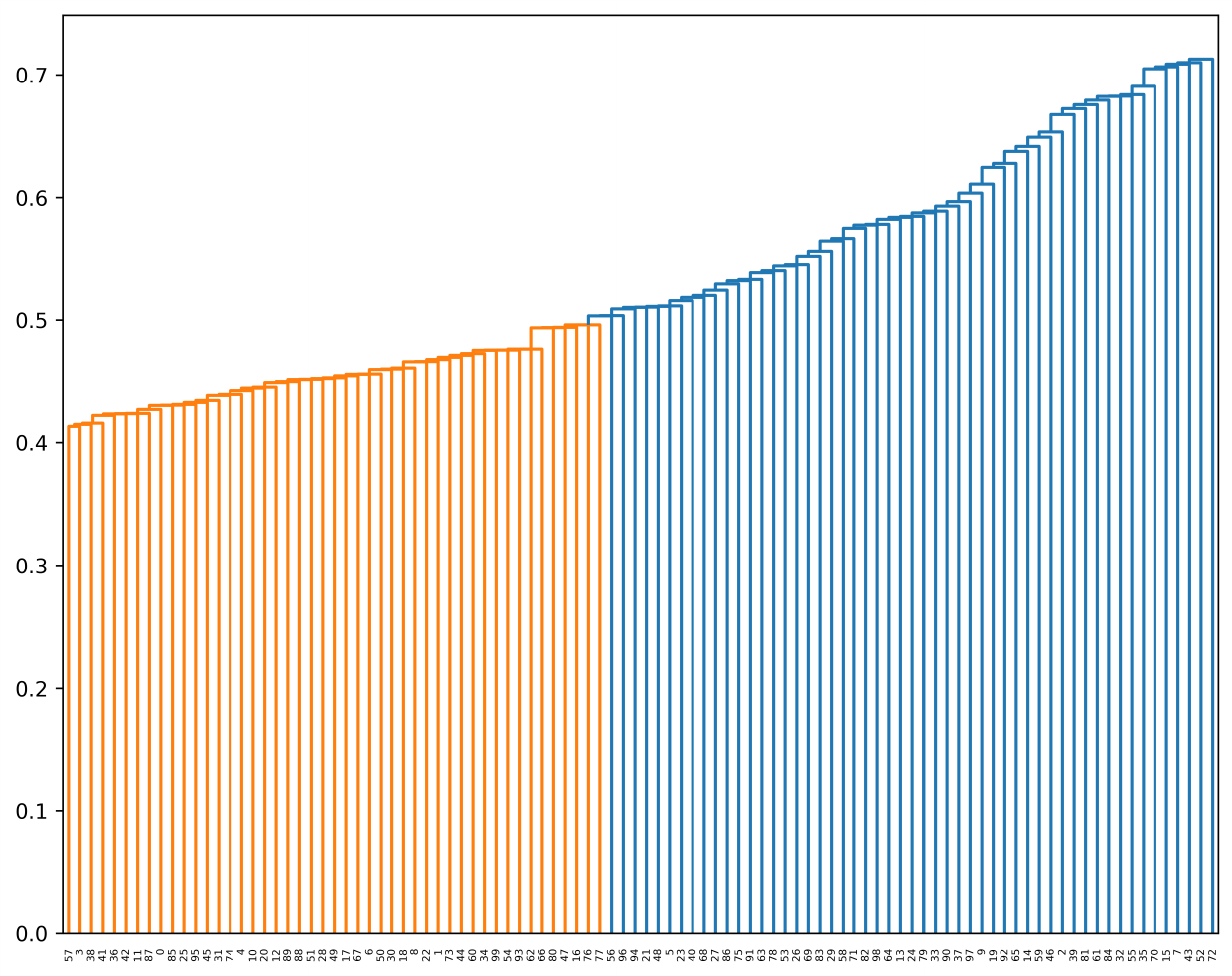}
        \caption{}
    \end{subfigure}\\
    \begin{subfigure}[b]{0.45\textwidth}
        \includegraphics[scale=0.25]{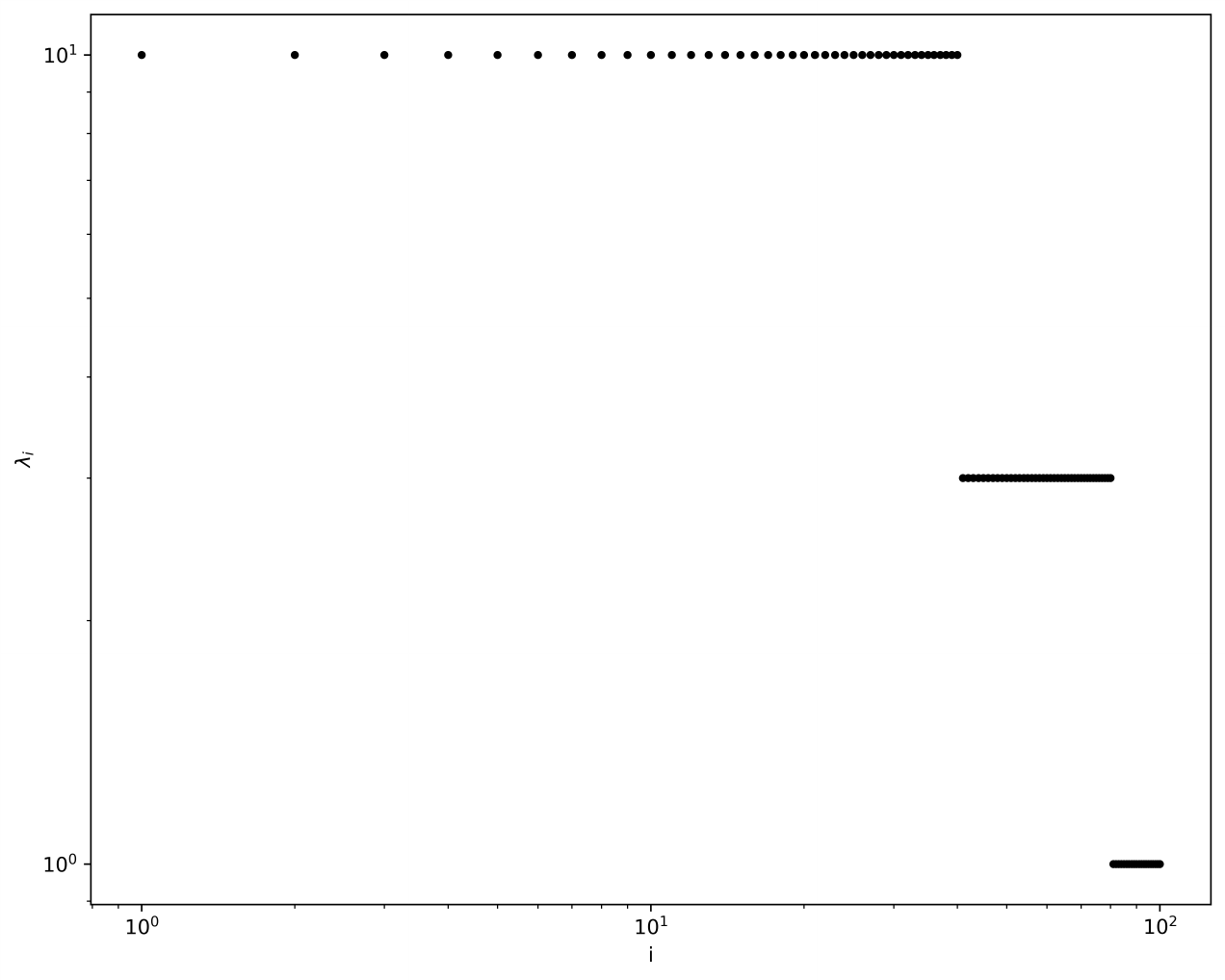}
        \caption{}
    \end{subfigure}
    \begin{subfigure}[b]{0.45\textwidth}
        \includegraphics[scale=0.25]{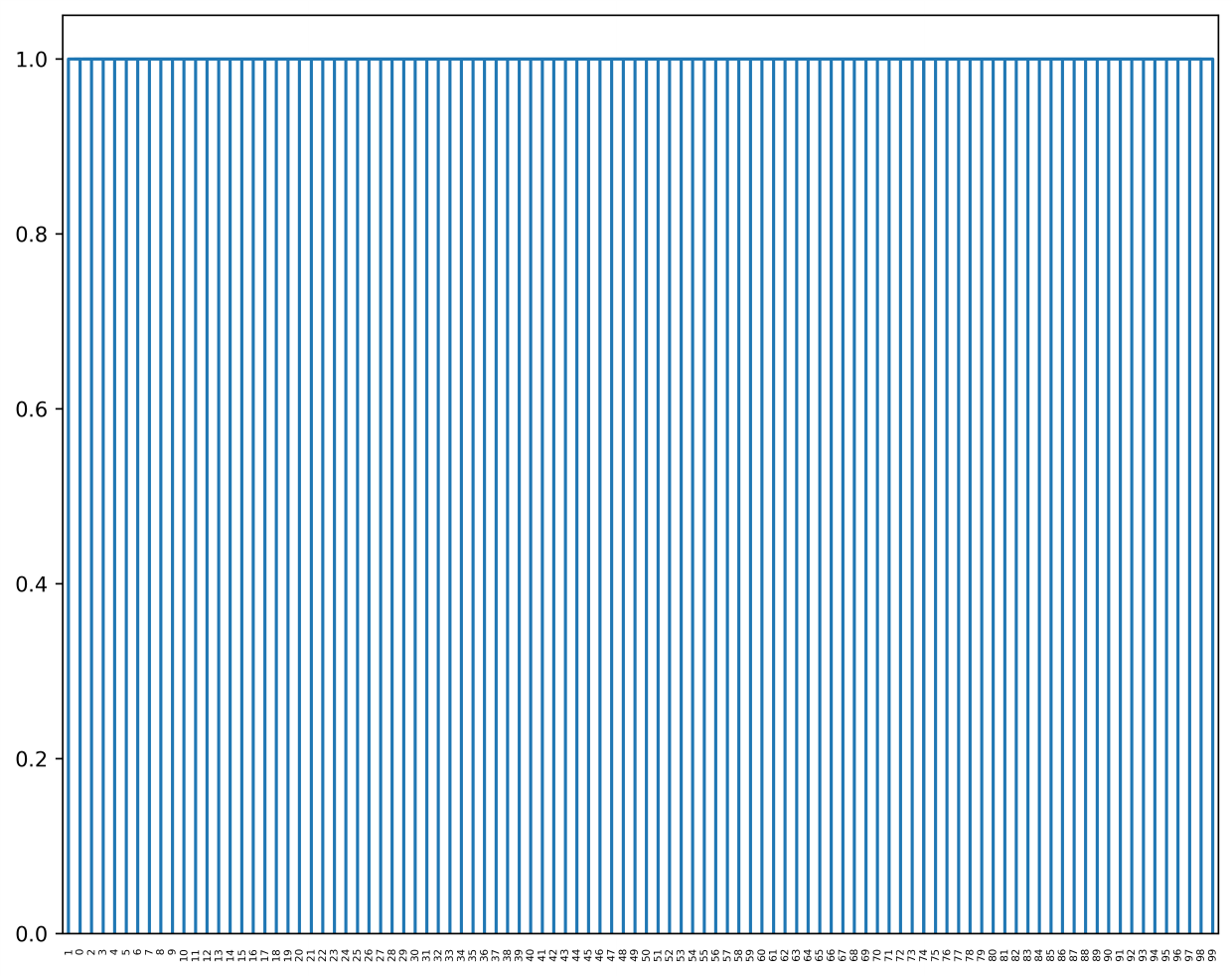}
        \caption{}
    \end{subfigure}\\
    \caption{Scree plot of ordered eigenvalues for model 1 (a), model 2 (c), and model 3 (e). Dendrogram under the same methodology of the hierarchical estimator with single linkage (SLCA) for model 1 (b), model 2 (d), and model 3 (f).  We set $p=100$, and $n=200$. For illustrative purpose the descendant links below a cluster node $k$ are equally colored if $k$ is the first node below the cut threshold $t(=0.7)$}
    \label{fig1}
\end{figure}

We consider the following covariance population models $\mathbf{\Sigma}$ and data generating processes $\mathbf{X}$
\begin{itemize}

\item[(1)] A completely nested hierarchical covariance model of the form:
    \begin{equation}
    \mathbf{\Sigma} = \mathbf{LL}^T
    \end{equation}
    where the matrix $\mathbf{L}$ of dimension $p\times p$ is given by
    \begin{equation}
    \mathbf{L} = \begin{pmatrix}
        \gamma & \gamma & \dots & \gamma & \gamma \\
        \gamma & \gamma & \dots &  \gamma & 0 \\
        \vdots &  \vdots & \ddots & \vdots & \vdots \\
        \gamma &  \gamma & \dots & 0 & 0 \\
        \gamma &  0 & \dots & 0 & 0 \\
    \end{pmatrix}
\end{equation}
setting $\gamma=0.1$ and $X_{ij}\sim \mathcal{N}(0,1)$, that is, each element $X_{ij}$ follows a standard Gaussian distribution.

Model 1 has the intriguing characteristic that their scree plot of eigenvalues follows a power law (see fig.\ref{fig1}a). We conjecture that it can capture the stylized fact of the complex interactions of the financial markets because is nested in all its levels (see dendrogram of figs.\ref{fig1}b). The population eigenvalues of this system are given by the solution of a tridiagonal symmetric Toeplitz matrix (see appendix~\ref{appendix}) having deep connections with Fibonacci and Lucas numbers~\cite{cahill2004fibonacci}.

\item[(2)] The one factor model studied in\cite{yang2015robust}
\begin{equation}
\mathbf{\Sigma} = \sigma^2\mathbf{bb}^T +\sigma_r^2\mathbf{I} 
\end{equation}
where the factor loading $\mathbf{b}\in \mathbb{R}^p\sim U(0.5,1.5)$, and the systematic and idiosyncratic variances are set to $\sigma=0.16$, $\sigma_r=0.2$, respectively. Here $\mathbf{X}$ follows a multivariate Student-T distribution with $d=3$ degrees of freedom. In model 2 only one large eigenvalue emerges and the rest are degenerate (see fig.\ref{fig1}c). The corresponding dendrogram shows a blurry hierarchy between the elements induced by the factor and noise~(see fig.\ref{fig1}d). This model is used in~\cite{yang2015robust} to test their deterministic equivalent-based estimator.

\item[(3)] A diagonal model where the population eigenvalues are distributed as follows: 20\% are equal to 1, 40\% are equal to 3, and 40\% are equal to 10. As in model 1,   $X_{ij}\sim \mathcal{N}(0,1)$. Here there are three distinct eigenvalues by construction: 10,3, and 1 (see fig.\ref{fig1}e). The dendrogram does not detect any cluster given the off-diagonal elements of the covariance matrix are zero(see fig.\ref{fig1}f). Model 3 is proposed originally in\cite{bai1998no} and used in \cite{ledoit2021shrinkage} to test their non-linear shrinkage estimators variants.
\end{itemize}

\section{Performance Metrics}
To measure the capital diversification in the estimated portfolio it is considered the Herfindahl–Hirschman index~(HHI) defined as~\cite{roncalli2013introduction}:
\begin{equation}
    \mathcal{HHI}(\mathbf{w}) = \sum_{i=1}^p w_i^2 
\end{equation}
where $w_i$, is the $i$-th allocation weight. Maximum diversification or perfect competition is reached when $HHI=1/p$, while a minimum diversification or monopoly has a value of $HHI=1$.

Another useful measure of diversification is the leverage defined as~\cite{roncalli2013introduction}
\begin{equation}
    \mathcal{L}(\mathbf{w}) = \sum_{i=1}^p |w_i| 
\end{equation}
This quantity reaches its minimum at $\mathcal{L}=1$ when do not exist short positions on the portfolio, while its maximum is unbounded and represents the multiplicative factor of the original capital to be borrowed to carry out the investment strategy. Usually, the restriction $\mathcal{L}=1$ is imposed on the objective function because the risk increases drastically with the growth of $\mathcal{L}$.

A measure of diversification in terms of risk is given by the Risk Diversification Index~(RDI)\cite{roncalli2013introduction, tasche2008capital}
\begin{equation}
\mathcal{RDI}(\mathbf{w}) = \frac{\sqrt{\mathbf{w}^T\mathbf{\Sigma w}}}{\mathbf{w}^T \sqrt{diag(\mathbf{\Sigma})} }
\end{equation}
This measure computes the ratio between the portfolio risk and the individual risk. In other words, the ratio between the systematic risk and the idiosyncratic risk. As such a $\mathcal{RDI}(\mathbf{w})<1$ represent a reduction in the investment risk by the optimization strategy, while a value of $\mathcal{RDI}(\mathbf{w})\geq 1$ is interpreted as an undesired behavior. 

Finally, a pure risk measure is given by the realized risk defined in  Eq.\ref{Risk_sample} as $\mathcal{R}_{out}^2$, where a baseline level is obtained by  $\mathcal{R}_{true}^2$~(see Eq.\ref{Risk_true}).

\section{Simulations}

We are interested in knowing the effect of the covariance estimators on the minimum variance portfolio under sample realization of data with population covariance matrix given by the models discussed in section~\ref{covariance_models}. 
Thus, numerical simulations were performed to compute the sample covariance matrix $\mathbf{S}$ applying each of the estimators described in Section~\ref{estimators}.

\begin{figure}[hbtp]
    \centering
    \begin{subfigure}[b]{0.45\textwidth}
        \includegraphics[scale=0.25]{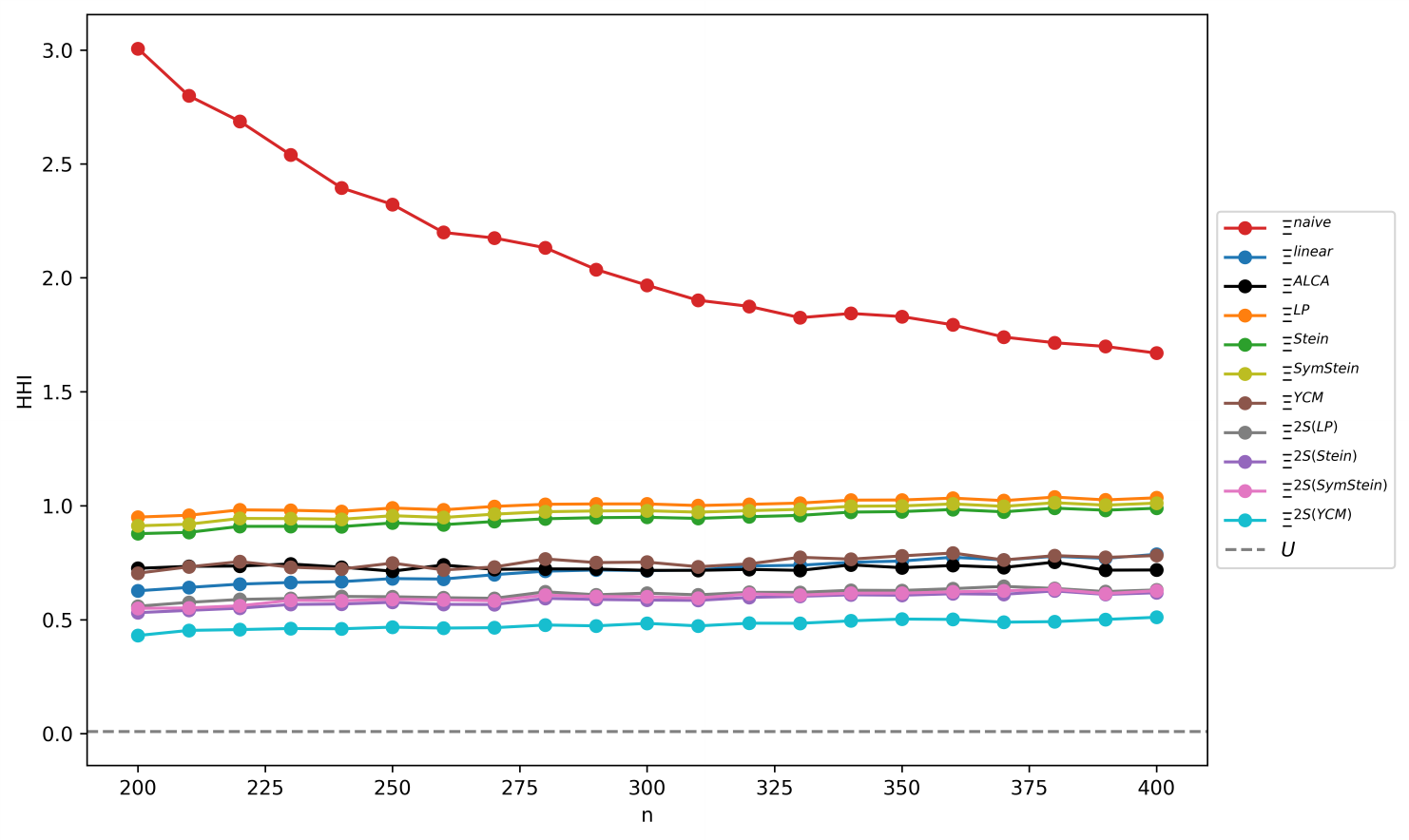}
        \caption{}
    \end{subfigure}
    \begin{subfigure}[b]{0.45\textwidth}
        \includegraphics[scale=0.25]{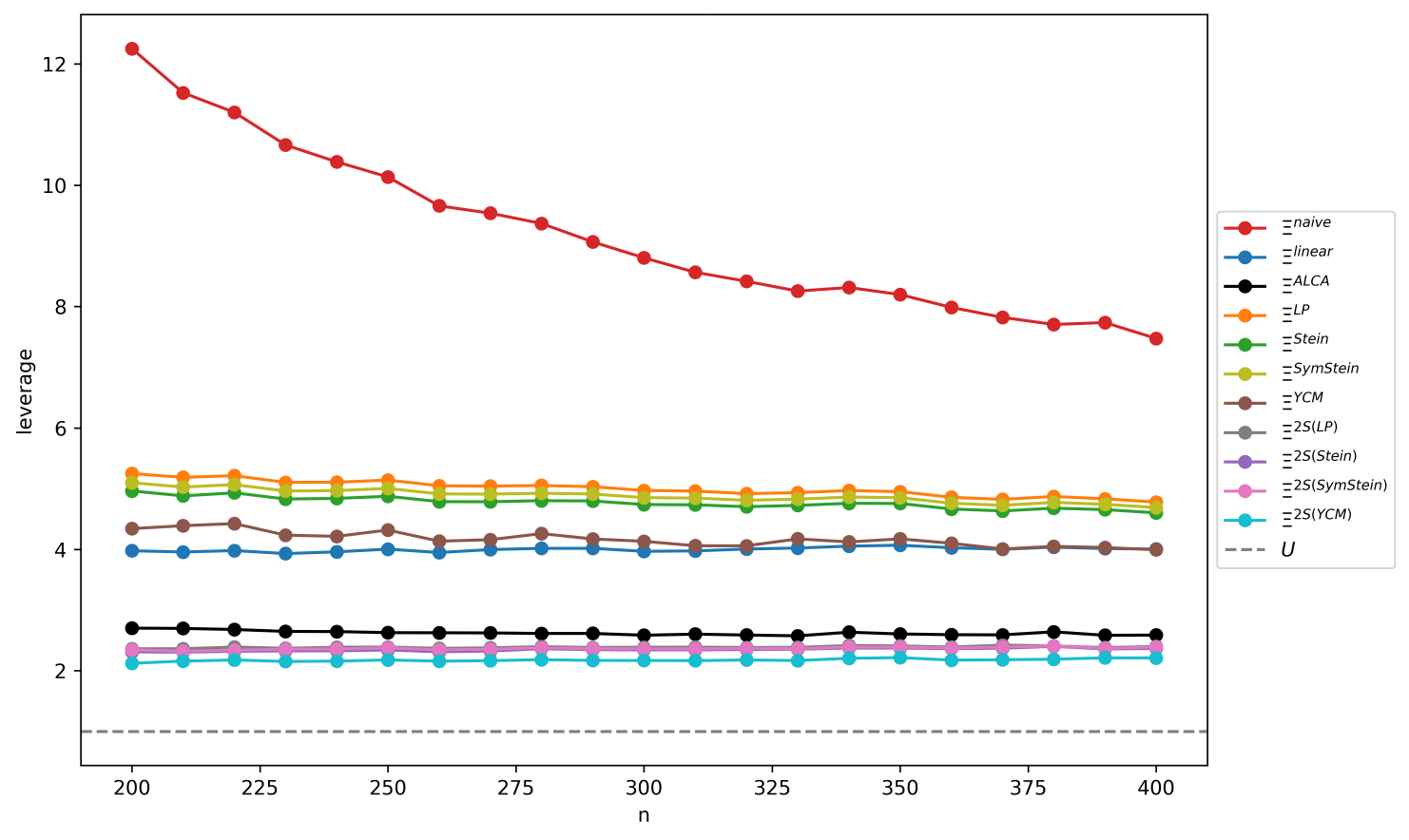}
        \caption{}
    \end{subfigure}\\
    \begin{subfigure}[b]{0.45\textwidth}
        \includegraphics[scale=0.25]{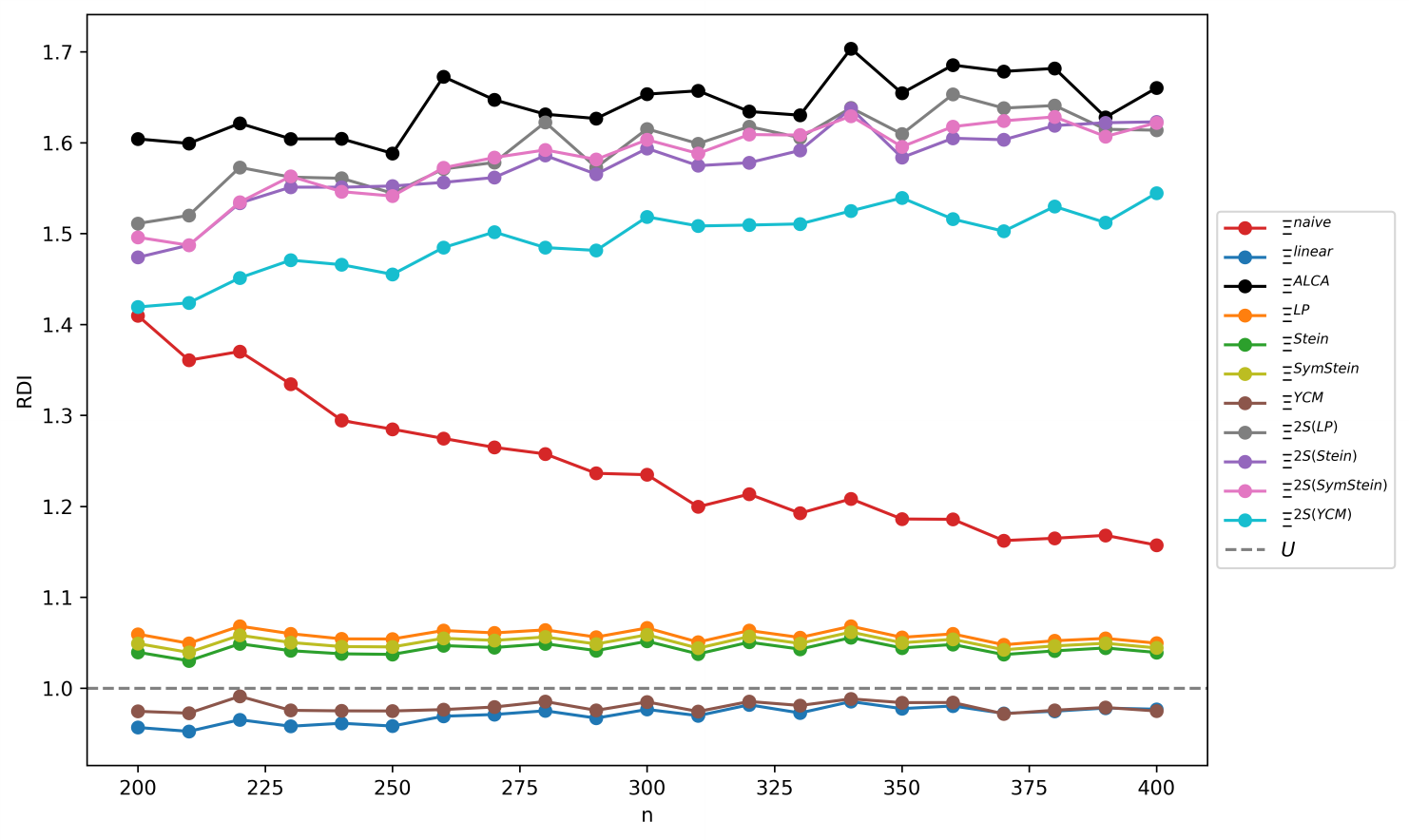}
        \caption{}
    \end{subfigure}
    \begin{subfigure}[b]{0.45\textwidth}
        \includegraphics[scale=0.25]{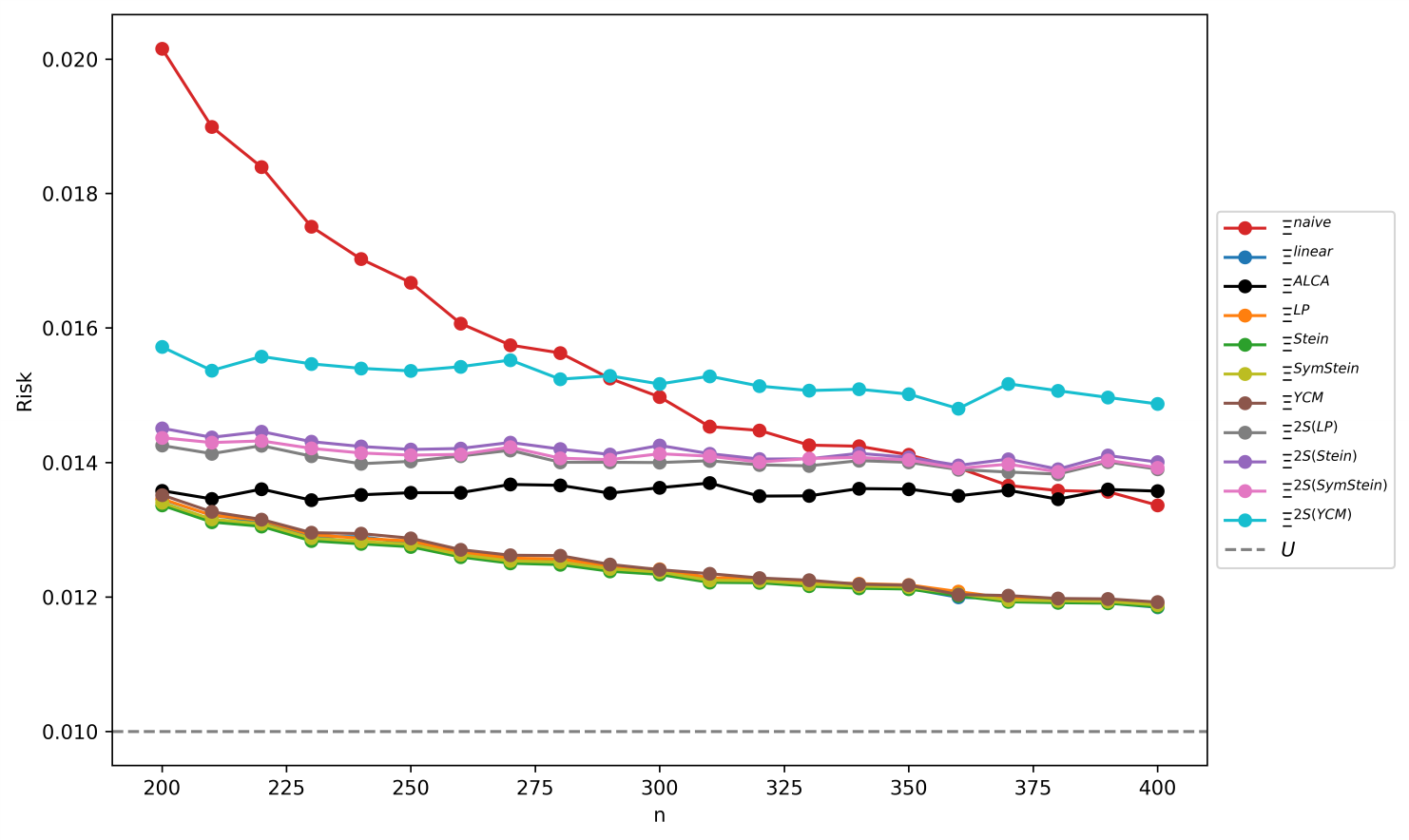}
        \caption{}
    \end{subfigure}\\
    \caption{Model 1. (a) Average $\mathcal{HHI}$, (b) Average $\mathcal{L}$, (c) Average $\mathcal{RDI}$, and (d) Average $\mathcal{R}^2_{out}$ as a function of the number of observations $n$ for each estimator of the covariance matrix. The uniform portfolio is superimposed as a reference baseline and denoted by the letter $U$ in the legend. The average is over $m=100$ realizations. The random matrix samples have dimensions $p=100$, and varying  $n\in[200, 400]$ with steps of $\Delta n = 10$. The optimal $\hat{\rho}$ of Eq.\ref{optimal_rho} is estimated by a grid search of size $20$.}
    \label{fig2}
\end{figure}

Fig.~\ref{fig2} shows $\mathcal{HHI}$(a), $\mathcal{L}$(b), $\mathcal{RDI}$(c), and $\mathcal{R}^2_{out}$(d) averaged over $m=100$ realizations of model 1, where the number of variables in the random matrix data $\mathbf{X}$ is set to $p=100$, and the number of observations has a variable dimension $n\in[200, 400]$ with steps $\Delta n = 10$. 
Here, the optimal $\hat{\rho}$ of Eq.\ref{optimal_rho} is estimated by a grid search of size $20$, and the average is taken over the metrics after optimization to have a consistent procedure for all estimators.

We can see that the best performance in terms of diversification and leverage is consistently obtained with the estimator $\mathbf{\Xi}^{2S(YCM)}$ for all $n$. On the other hand, $\mathcal{RDI}<1$ is only achieved for $\mathbf{\Xi}^{linear}$ and $\mathbf{\Xi}^{YCM}$, where the first estimator performs slightly better as $n\rightarrow p$. Thus, it may be more convenient in extremely high-dimensional situations.  The uniform portfolio gets the value  $\mathcal{RDI}=1$. Thus, it does not get a good risk diversification under this complex model. For $\mathcal{R}^2_{out}$, the performance is similar for estimators based on the variants of the non-linear shrinkage formula: $\mathbf{\Xi}^{LP}$, $\mathbf{\Xi}^{Stein}$, $\mathbf{\Xi}^{SymStein}$; and the proposal built with equivalent deterministic $\mathbf{\Xi}^{YCM}$.

\begin{figure}[hbtp]
    \centering
    \begin{subfigure}[b]{0.45\textwidth}
        \includegraphics[scale=0.25]{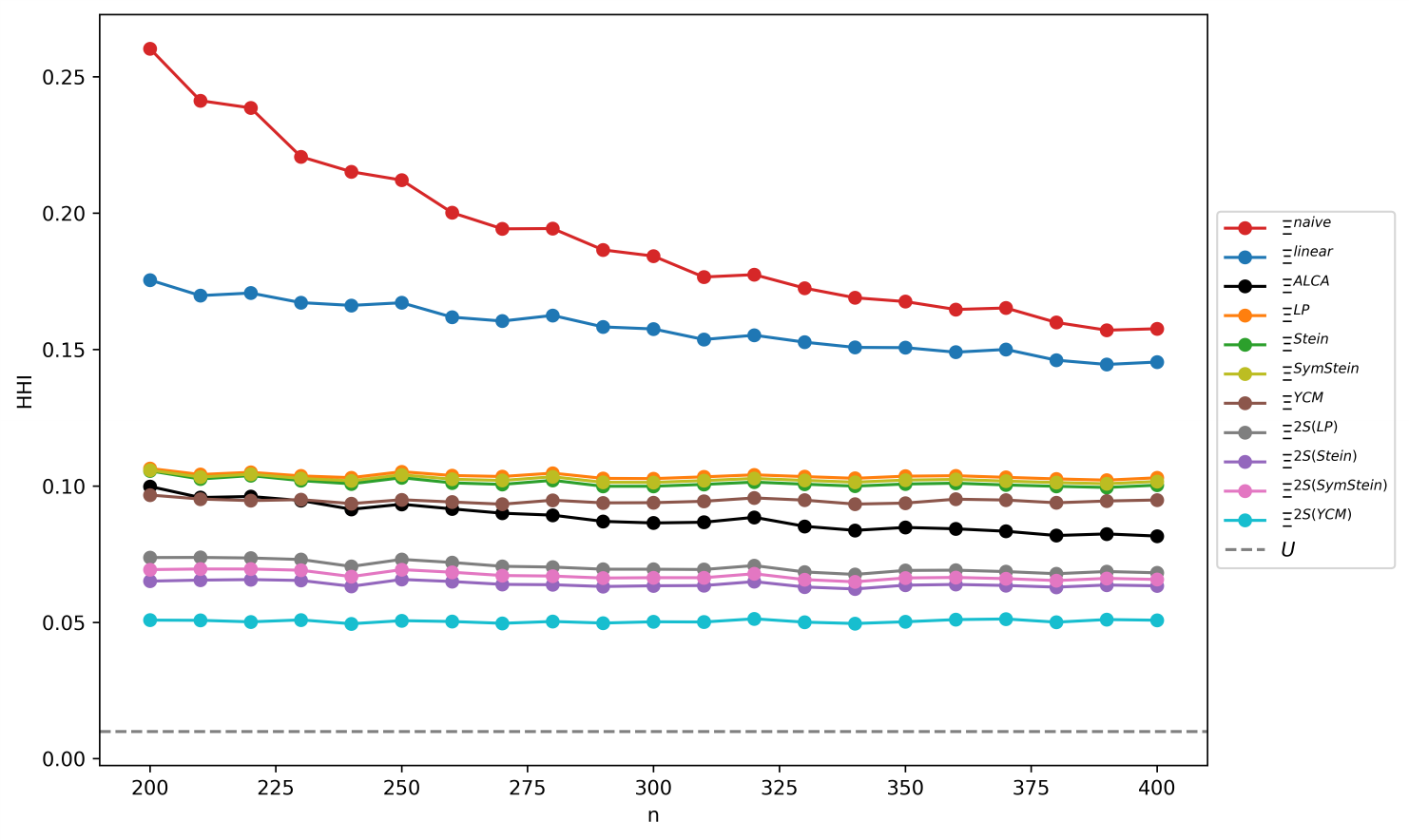}
        \caption{}
    \end{subfigure}
    \begin{subfigure}[b]{0.45\textwidth}
        \includegraphics[scale=0.25]{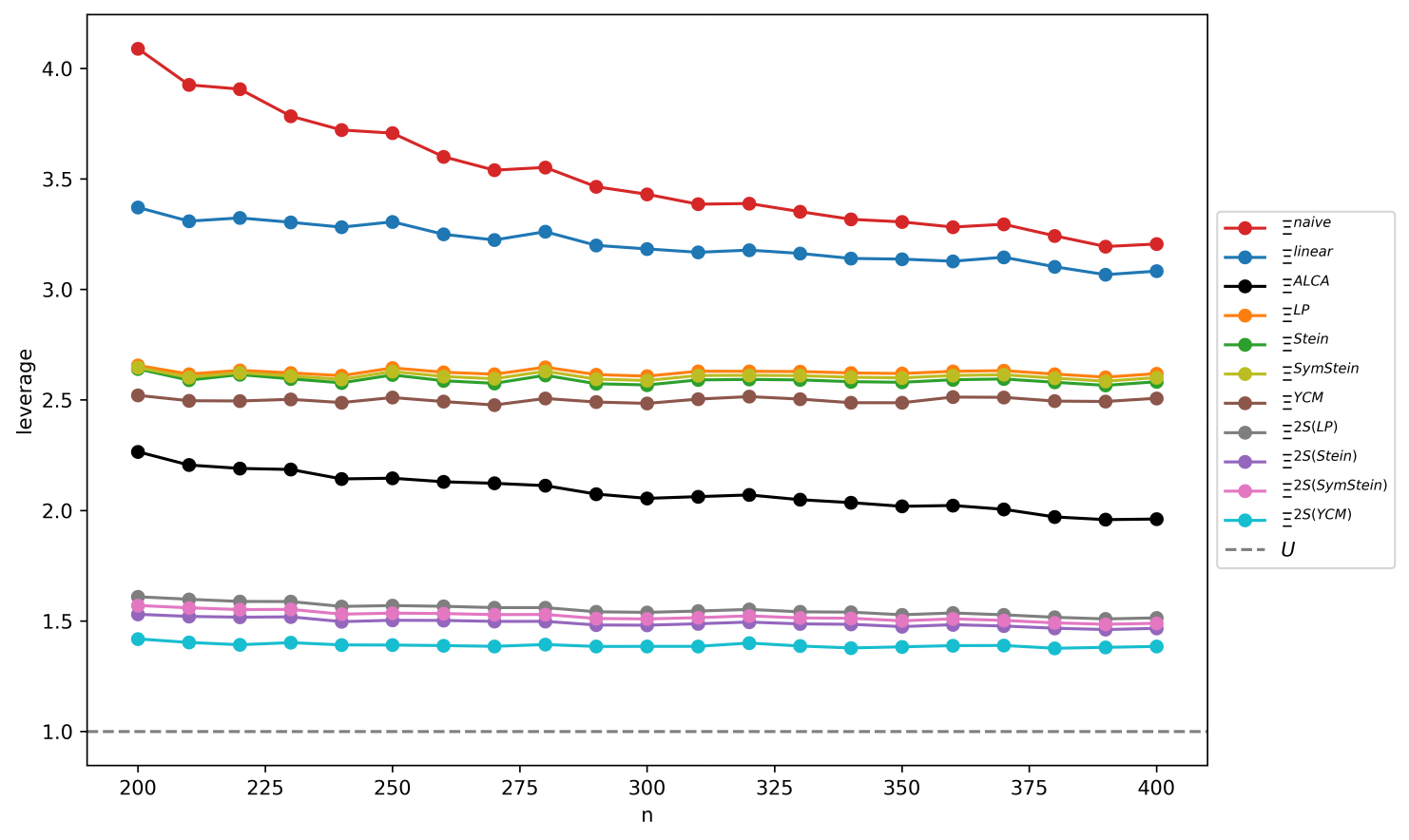}
        \caption{}
    \end{subfigure}\\
    \begin{subfigure}[b]{0.45\textwidth}
        \includegraphics[scale=0.25]{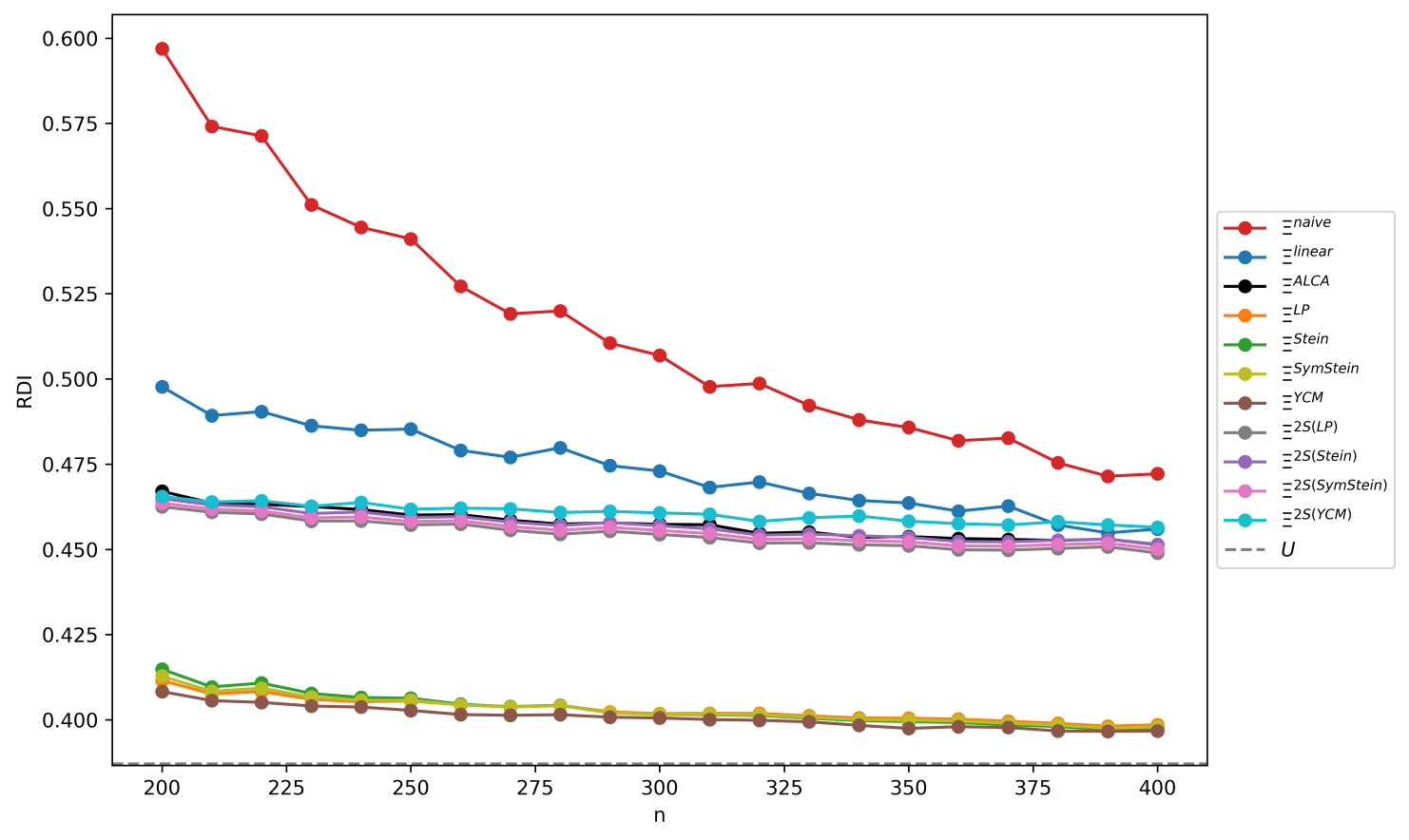}
        \caption{}
    \end{subfigure}
    \begin{subfigure}[b]{0.45\textwidth}
        \includegraphics[scale=0.25]{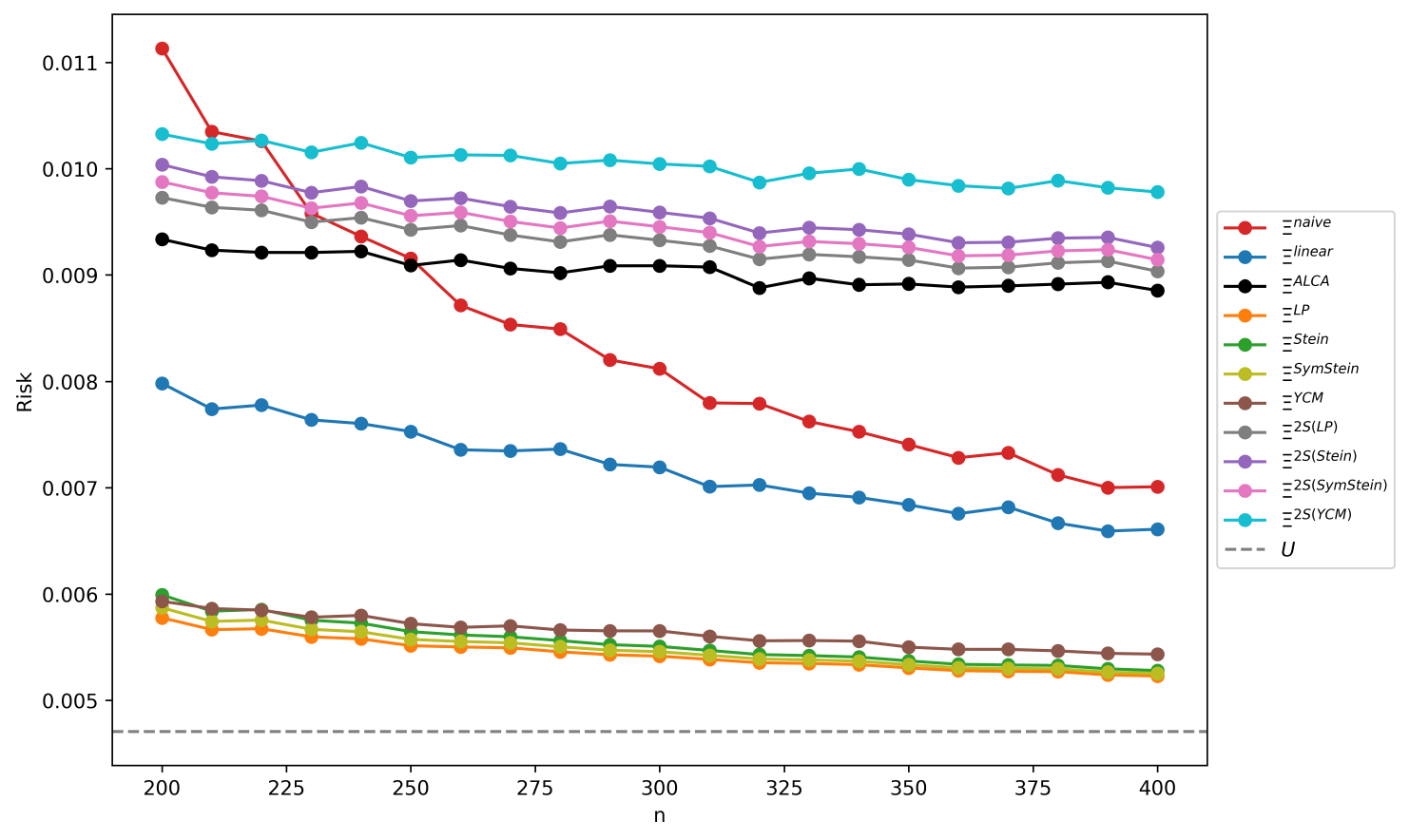}
        \caption{}
    \end{subfigure}\\
    \caption{Model 2. (a) Average $\mathcal{HHI}$, (b) Average $\mathcal{L}$, (c) Average $\mathcal{RDI}$, and (d) Average $\mathcal{R}^2_{out}$ as a function of the number of observations $n$ for each estimator of the covariance matrix. The uniform portfolio is superimposed as a reference baseline and denoted by the letter $U$ in the legend. The average is over $m=100$ realizations. The random matrix samples have dimensions $p=100$, and varying  $n\in[200, 400]$ with steps of $\Delta n = 10$. The optimal $\hat{\rho}$ of Eq.\ref{optimal_rho} is estimated by a grid search of size $20$.}
    \label{fig3}
\end{figure}

Fig.\ref{fig3} shows equivalent graphs as in \ref{fig2} but for model 2. Likewise, for $\mathcal{HHI}$ and $\mathcal{L}$ we can observe a similar trend in the performance of the two-step estimators, except that smaller values are reached in magnitude. In particular, the values of the top estimator $\mathbf{\Xi}^{2S(YCM)}$ are one order of magnitude smaller in the metric $\mathcal{HHI}$ for this single-factor model with respect to the hierarchically nested model.
On the other hand, it is found that $\mathbf{\Xi}^{YCM}$ and the non-linear estimators minimize $\mathcal{RDI}$ and $\mathcal{R}^2_{out}$. In the first case $\mathbf{\Xi}^{YCM}$ obtains the best performance, while for the realized risk, $\mathbf{\Xi}^{LP}$ reach the minimum values for the range of $n$.
In this case, it is interesting to observe that $\mathcal{RDI}<1$ for all estimators.

\begin{figure}[hbtp]
    \centering
    \begin{subfigure}[b]{0.45\textwidth}
        \includegraphics[scale=0.25]{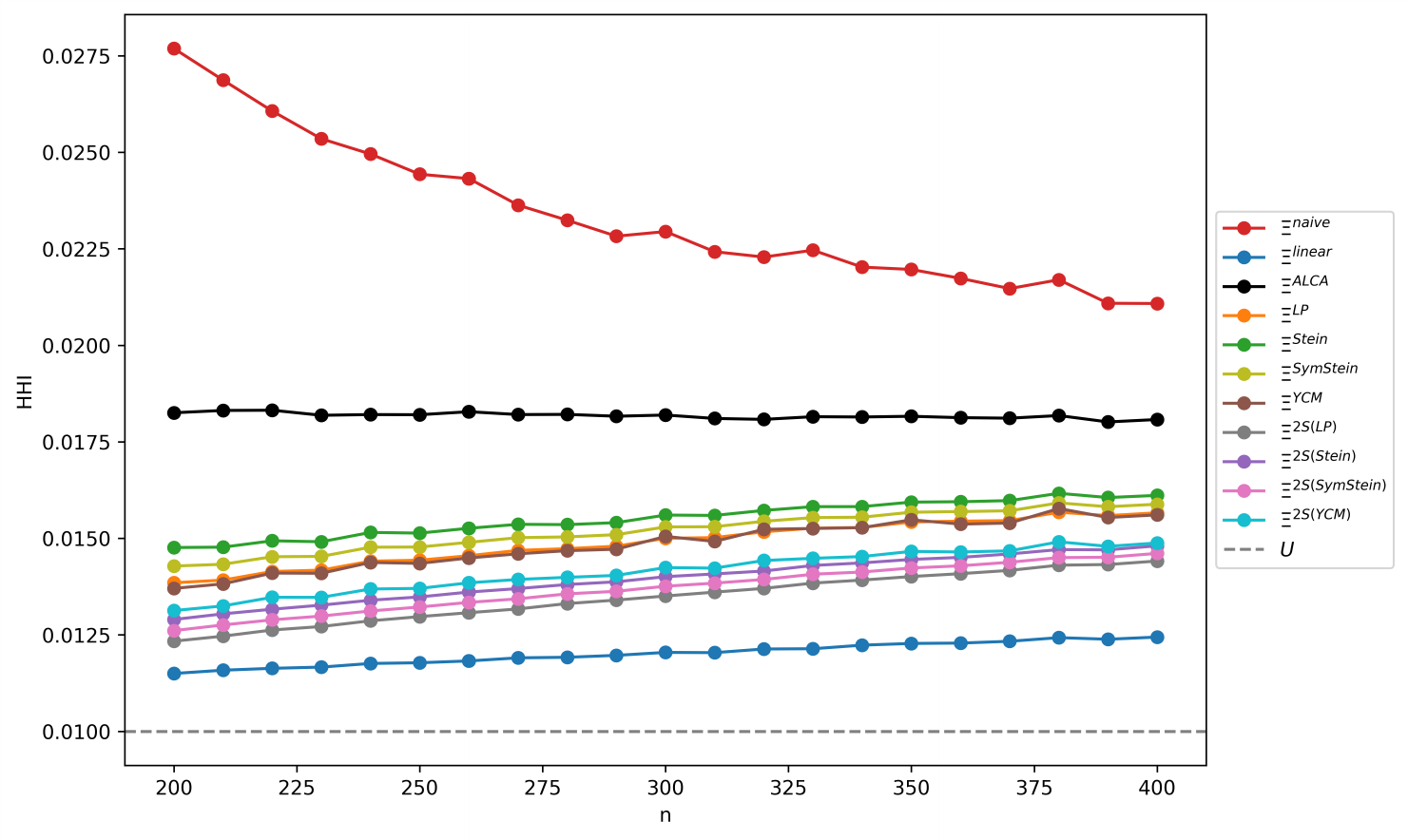}
        \caption{}
    \end{subfigure}
    \begin{subfigure}[b]{0.45\textwidth}
        \includegraphics[scale=0.25]{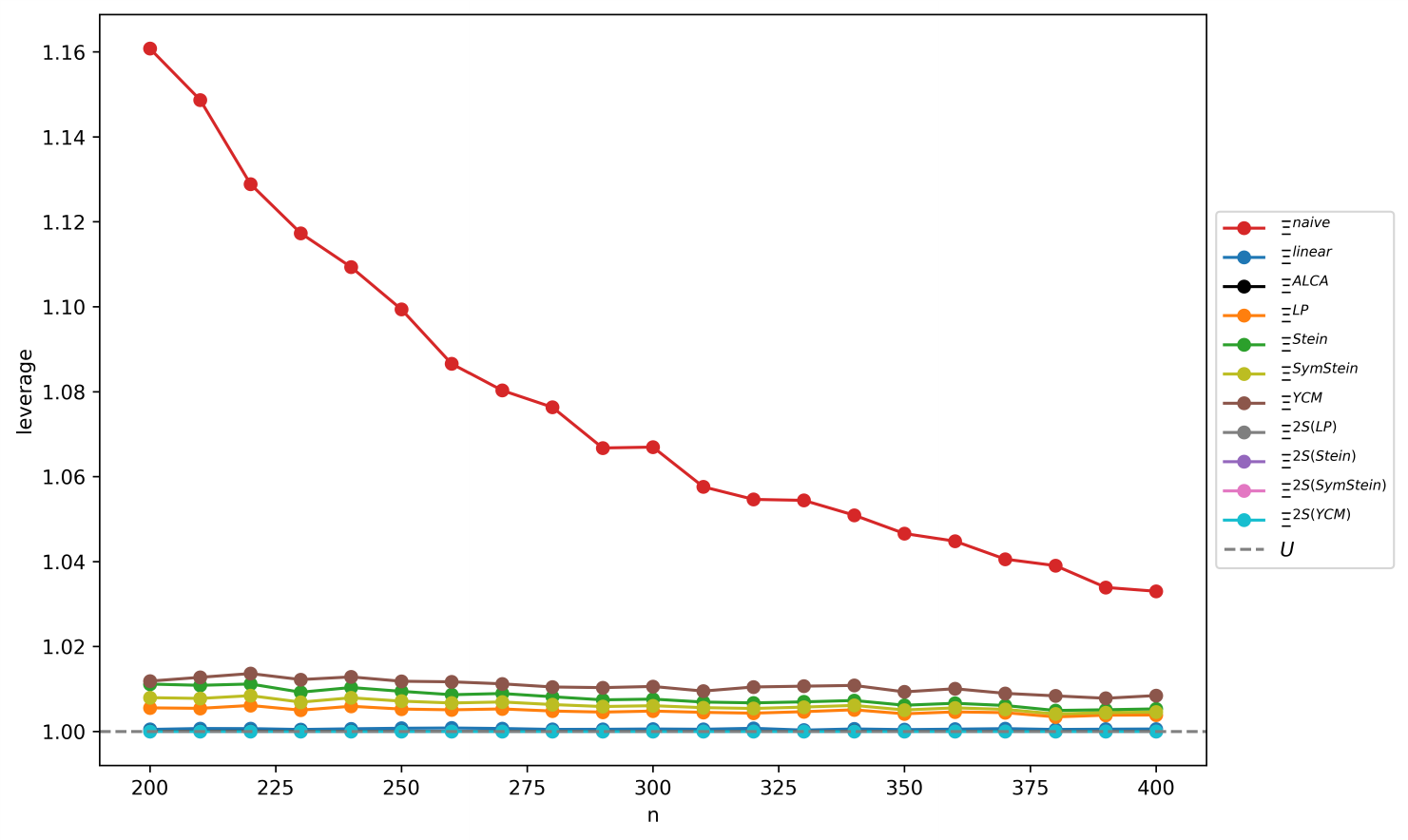}
        \caption{}
    \end{subfigure}\\
    \begin{subfigure}[b]{0.45\textwidth}
        \includegraphics[scale=0.25]{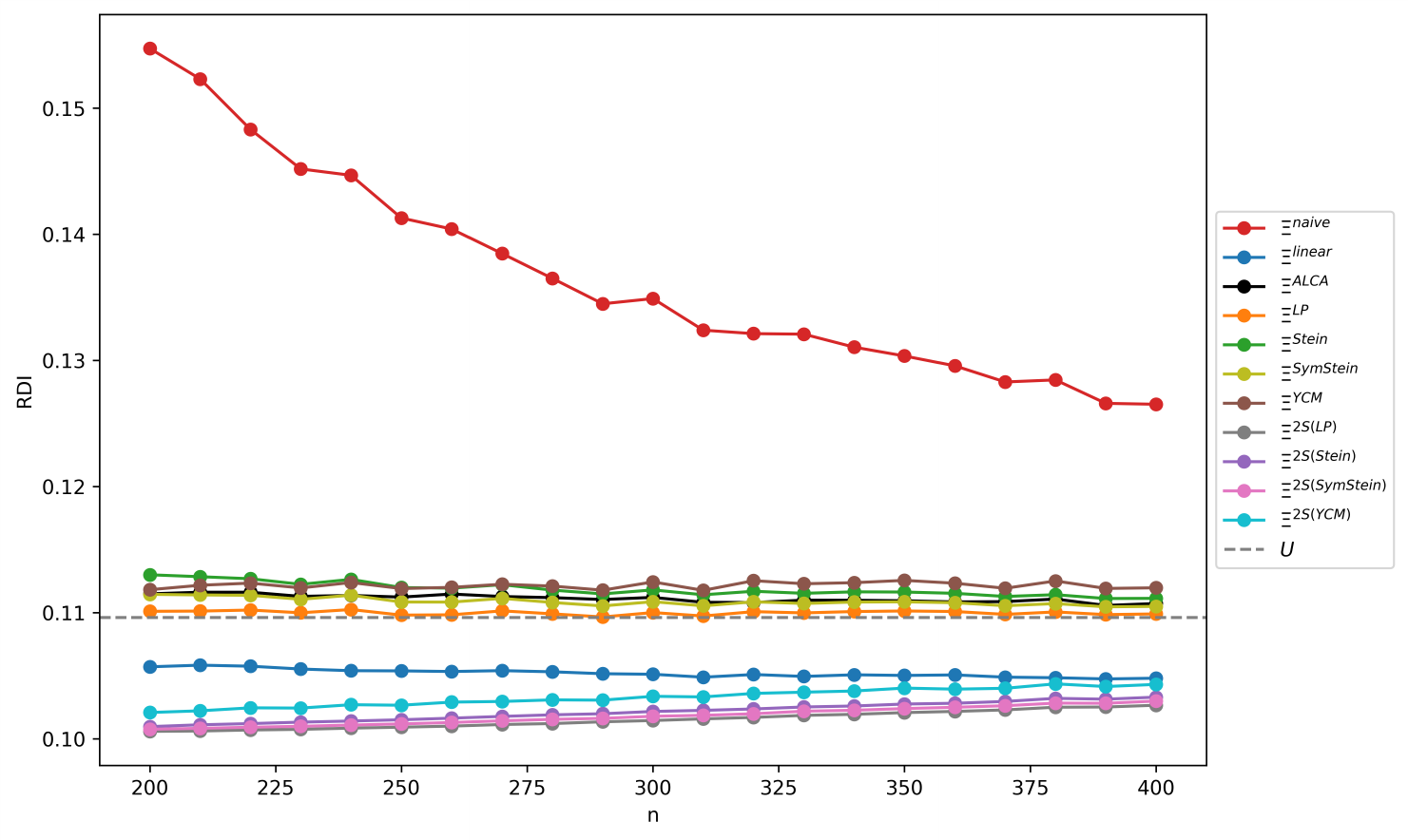}
        \caption{}
    \end{subfigure}
    \begin{subfigure}[b]{0.45\textwidth}
        \includegraphics[scale=0.25]{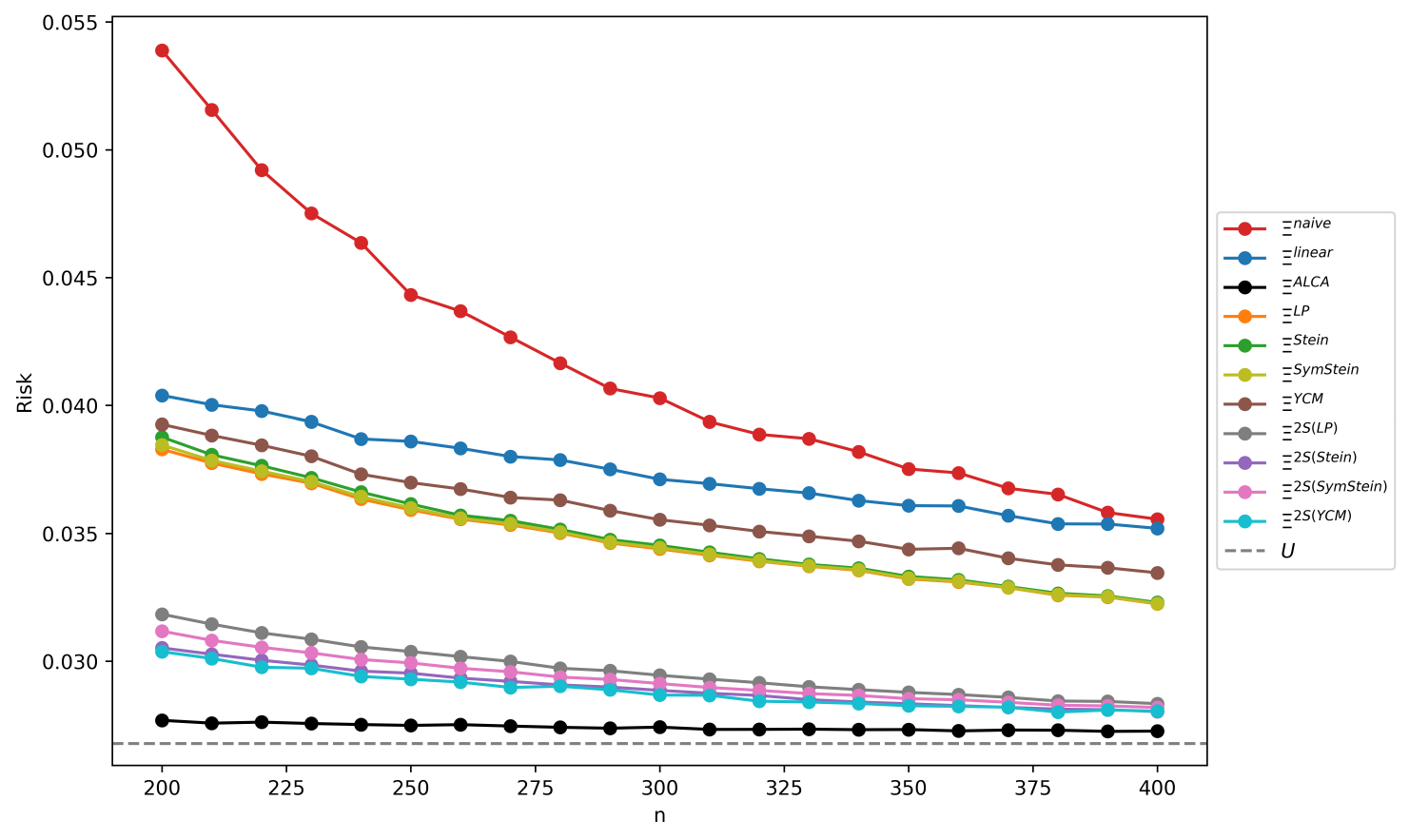}
        \caption{}
    \end{subfigure}\\
    \caption{Model 3. (a) Average $\mathcal{HHI}$, (b) Average $\mathcal{L}$, (c) Average $\mathcal{RDI}$, and (d) Average $\mathcal{R}^2_{out}$ as a function of the number of observations $n$ for each estimator of the covariance matrix. The uniform portfolio is superimposed as a reference baseline and denoted by the letter $U$ in the legend. The average is over $m=100$ realizations. The random matrix samples have dimensions $p=100$, and varying  $n\in[200, 400]$ with steps of $\Delta n = 10$. The optimal $\hat{\rho}$ of Eq.\ref{optimal_rho} is estimated by a grid search of size $20$.}
    \label{fig4}
\end{figure}

For model 3, the equivalent graphs can be seen in Fig.\ref{fig4}.
In this diagonal model, the best performance in terms of $\mathcal{HHI}$ is obtained by $\mathbf{\Xi}^{linear}$. For leverage, the two-step estimators and $\mathbf{\Xi}^{linear}$ reach the minimum bound $\mathcal{L}=1$, i.e., the case where there are no present short positions in the allocation weights.
Similarly, for $\mathcal{RDI}$ the best results are obtained for the two-step estimators and the linear estimator, in fact improving the performance of the uniform portfolio. Finally, the two-step estimators are outperformed by $\mathbf{\Xi}^{ALCA}$ for the realized risk ($\mathcal{R}^2_{out}$), obtaining a stable value for all $n$.

We are also interested in knowing the effect of the estimators on the different capital allocation strategies presented in section~\ref{allocation_models}. For the sake of comparison and to reduce computational demand, we have restricted ourselves to the case $p=100, n=200$ to compute the data sample $\mathbf{X}$.
In tables~\ref{table1}, ~\ref{table2}, and \ref{table3}, one can see the performance of the metrics for an average of $m=1000$ realizations of the population covariance matrix given by the models discussed in section~\ref{covariance_models}. In our simulation framework the in-sample covariance matrix $\mathbf{S}_{in}$ is given by a simple realization of $\mathbf{\Sigma}$ and then it is filtered using the estimators described in Section~\ref{estimators}. Whereas $\mathbf{S}_{out}$ is considered the population $\mathbf{\Sigma}$.
As a reference value, we have included the baseline case $\mathbf{S}_{in}=\mathbf{\Sigma}$ on the tables.

In table\ref{table1} we can see that the best performance in terms of $\mathcal{HHI}$ is achieved with the estimator $\Xi^{2S(YCM)}$ for models 1 and 2, regardless of the optimization strategy.
For model 3 the estimator $\Xi^{linear}$ consistently achieves the lowest values for the MVP, MVP+, and HRP strategies.
It is interesting to note that $\mathcal{HHI}=1$ in model 1, both for the MVP and MVP+ strategies, i.e., allocations are localized in one of the assets. Therefore, in this complex structure model, it is particularly important to apply an estimator to increase diversification.
It is also important to mention that the HRP strategy systematically increases the diversification (low values of $\mathcal{HHI}$) for the three models regardless of the estimator.
\begin{table}[htbp]
\tiny
\centering
\caption{Average $\mathcal{HHI}$ over $m=1000$ realizations of model 1, model 2, and model 3 on the allocation strategies MVP, MPV+, and HRP. The random matrix samples have dimensions $p=100, n=200$. }
\begin{tabular}{lrrrrrrrrr}
\hline
 & \multicolumn{ 3}{l|}{Model 1} & \multicolumn{ 3}{l|}{Model 2} & \multicolumn{ 3}{l|}{Model 3} \\ \hline
estimator & \multicolumn{1}{l|}{MVP} & \multicolumn{1}{l|}{MVP+} & \multicolumn{1}{l|}{HRP} & \multicolumn{1}{l|}{MVP} & \multicolumn{1}{l|}{MVP+} & \multicolumn{1}{l|}{HRP} & \multicolumn{1}{l|}{MVP} & \multicolumn{1}{l|}{MVP+} & \multicolumn{1}{l|}{HRP} \\ \hline
$\Sigma$ & 1.0000 & 1.0000 & 0.1367 & 0.1004 & 0.0620 & 0.0141 & 0.0178 & 0.0178 & 0.0178 \\ \hline
$\Xi^{naive}$ & 3.0121 & 0.9215 & 0.1430 & 0.2583 & 0.0821 & 0.0149 & 0.0279 & 0.0229 & 0.0187 \\ \hline
$\Xi^{linear}$ & 0.6297 & 0.5272 & 0.0964 & 0.1731 & 0.0764 & 0.0147 & $\mathbf{0.0115}$ & $\mathbf{0.0115}$ & $\mathbf{0.0114}$ \\ \hline
$\Xi^{ALCA}$ & 0.7336 & 0.7326 & 0.1367 & 0.0990 & 0.0729 & 0.0136 & 0.0183 & 0.0183 & 0.0184 \\ \hline
$\Xi^{LP}$ & 0.9577 & 0.7508 & 0.1284 & 0.1052 & 0.0730 & 0.0149 & 0.0138 & 0.0138 & 0.0125 \\ \hline
$\Xi^{Stein}$ & 0.8826 & 0.7044 & 0.1231 & 0.1043 & 0.0657 & 0.0146 & 0.0147 & 0.0146 & 0.0131 \\ \hline
$\Xi^{SymStein}$ & 0.9183 & 0.7271 & 0.1257 & 0.1046 & 0.0692 & 0.0148 & 0.0143 & 0.0141 & 0.0128 \\ \hline
$\Xi^{YCM}$ & 0.7190 & 0.5716 & 0.1023 & 0.0950 & 0.0538 & 0.0139 & 0.0138 & 0.0136 & 0.0134 \\ \hline
$\Xi^{2S(LP)}$ & 0.5738 & 0.6028 & 0.1222 & 0.0743 & 0.0697 & 0.0137 & 0.0124 & 0.0124 & 0.0123 \\ \hline
$\Xi^{2S(Stein)}$ & 0.5404 & 0.5733 & 0.1162 & 0.0658 & 0.0615 & 0.0134 & 0.0129 & 0.0129 & 0.0129 \\ \hline
$\Xi^{2S(SymStein)}$ & 0.5567 & 0.5875 & 0.1194 & 0.0699 & 0.0655 & 0.0136 & 0.0126 & 0.0126 & 0.0126 \\ \hline
$\Xi^{2S(YCM)}$ & $\mathbf{0.4427}$ & $\mathbf{0.4759}$ & $\mathbf{0.0928}$ & $\mathbf{0.0509}$ & $\mathbf{0.0474}$ & $\mathbf{0.0127}$ & 0.0132 & 0.0132 & 0.0131 \\ \hline
\end{tabular}
\label{table1}
\end{table}

The results for $\mathcal{L}$ are shown in table~\ref{table2}.
In this case, the investment strategies MVP+ and HRP have been added simply for completeness. They always obtain the value of 1 by construction. The only strategy of interest here is MVP, where it can be seen that for models 1 and 2 the estimator $\Xi^{2S(YCM)}$ again obtains the best performance. In model 3, only the sample estimator obtains values significantly greater than 1. For this diagonal model, any estimator practically reduces the leverage to a minimum.
Note that the leverage of the population covariance matrix of model 2 is greater than 1 and that obtained by the estimator $\Xi^{2S(YCM)}$. Thus the estimator manages to reduce the baseline leverage.
\begin{table}[htbp]
\tiny
\centering
\caption{Average $\mathcal{L}$ over $m=1000$ realizations of model 1, model 2, and model 3 on the allocation strategies MVP, MPV+, and HRP. The random matrix samples have dimensions $p=100, n=200$. }
\begin{tabular}{lrrrrrrrrr}
\hline
 & \multicolumn{ 3}{l|}{Model 1} & \multicolumn{ 3}{l|}{Model 2} & \multicolumn{ 3}{l|}{Model 3} \\ \hline
estimator & \multicolumn{1}{l|}{MVP} & \multicolumn{1}{l|}{MVP+} & \multicolumn{1}{l|}{HRP} & \multicolumn{1}{l|}{MVP} & \multicolumn{1}{l|}{MVP+} & \multicolumn{1}{l|}{HRP} & \multicolumn{1}{l|}{MVP} & \multicolumn{1}{l|}{MVP+} & \multicolumn{1}{l|}{HRP} \\ \hline
$\Sigma$ & 1.00 & 1.00 & 1.00 & 2.63 & 1.00 & 1.00 & 1.00 & 1.00 & 1.00 \\ \hline
$\Xi^{naive}$ & 12.19 & 1.00 & 1.00 & 4.07 & 1.00 & 1.00 & 1.16 & 1.00 & 1.00 \\ \hline
$\Xi^{linear}$ & 3.96 & 1.00 & 1.00 & 3.35 & 1.00 & 1.00 & 1.00 & 1.00 & 1.00 \\ \hline
$\Xi^{ALCA}$ & 2.69 & 1.00 & 1.00 & 2.25 & 1.00 & 1.00 & 1.00 & 1.00 & 1.00 \\ \hline
$\Xi^{LP}$ & 5.24 & 1.00 & 1.00 & 2.63 & 1.00 & 1.00 & 1.01 & 1.00 & 1.00 \\ \hline
$\Xi^{Stein}$ & 4.95 & 1.00 & 1.00 & 2.62 & 1.00 & 1.00 & 1.01 & 1.00 & 1.00 \\ \hline
$\Xi^{SymStein}$ & 5.09 & 1.00 & 1.00 & 2.62 & 1.00 & 1.00 & 1.01 & 1.00 & 1.00 \\ \hline
$\Xi^{YCM}$ & 4.38 & 1.00 & 1.00 & 2.50 & 1.00 & 1.00 & 1.01 & 1.00 & 1.00 \\ \hline
$\Xi^{2S(LP)}$ & 2.38 & 1.00 & 1.00 & 1.61 & 1.00 & 1.00 & 1.00 & 1.00 & 1.00 \\ \hline
$\Xi^{2S(Stein)}$ & 2.31 & 1.00 & 1.00 & 1.54 & 1.00 & 1.00 & 1.00 & 1.00 & 1.00 \\ \hline
$\Xi^{2S(SymStein)}$ & 2.34 & 1.00 & 1.00 & 1.57 & 1.00 & 1.00 & 1.00 & 1.00 & 1.00 \\ \hline
$\Xi^{2S(YCM)}$ & $\mathbf{2.15}$ & 1.00 & 1.00 & $\mathbf{1.40}$ & 1.00 & 1.00 & 1.00 & 1.00 & 1.00 \\ \hline
\end{tabular}
\label{table2}
\end{table}

The behavior of the $\mathcal{RDI}$ can be reviewed in table~\ref{table3}. We can see a heterogeneous behavior in the performance depending on the model and the investment strategy. For model 1, the estimator $\Xi^{linear}$ obtains the best performance under MVP and MVP+. While $\Xi^{naive}$ is the best estimator in the HRP strategy. Therefore, in this case, applying an estimator does not improve risk diversification.
For model 2, the estimator $\Xi^{YCM}$ is the best under the MVP and MVP+ strategies, while $\Xi^{LP}$ has the best performance when applying HRP.
In model 3, it is observed that the estimator $\Xi^{2S(LP)}$ obtains the best performance systematically for the three investment strategies. In general model 3 reach lower values compared to models 1 and 2.
\begin{table}[htbp]
\tiny
\centering
\caption{Average $\mathcal{RDI}$ over $m=1000$ realizations of model 1, model 2, and model 3 on the allocation strategies MVP, MPV+, and HRP. The random matrix samples have dimensions $p=100, n=200$. }
\begin{tabular}{lrrrrrrrrr}
\hline
 & \multicolumn{ 3}{l|}{Model 1} & \multicolumn{ 3}{l|}{Model 2} & \multicolumn{ 3}{l|}{Model 3} \\ \hline
estimator & \multicolumn{1}{l|}{MVP} & \multicolumn{1}{l|}{MVP+} & \multicolumn{1}{l|}{HRP} & \multicolumn{1}{l|}{MVP} & \multicolumn{1}{l|}{MVP+} & \multicolumn{1}{l|}{HRP} & \multicolumn{1}{l|}{MVP} & \multicolumn{1}{l|}{MVP+} & \multicolumn{1}{l|}{HRP} \\ \hline
$\Sigma$ & 1.0000 & 1.0000 & 0.7190 & 0.3871 & 0.4685 & 0.5663 & 0.1096 & 0.1096 & 0.1096 \\ \hline
$\Xi^{naive}$ & 1.4274 & 0.9313 & $\mathbf{0.7184}$ & 0.5930 & 0.4969 & 0.5686 & 0.1553 & 0.1319 & 0.1122 \\ \hline
$\Xi^{linear}$ & $\mathbf{0.9650}$ & $\mathbf{0.8577}$ & 0.7295 & 0.4940 & 0.4937 & 0.5691 & 0.1058 & 0.1057 & 0.1009 \\ \hline
$\Xi^{ALCA}$ & 1.6230 & 0.9308 & 0.7203 & 0.4661 & 0.4954 & 0.5785 & 0.1117 & 0.1117 & 0.1110 \\ \hline
$\Xi^{LP}$ & 1.0679 & 0.9049 & 0.7208 & 0.4097 & 0.4897 & $\mathbf{0.5673}$ & 0.1100 & 0.1094 & 0.1008 \\ \hline
$\Xi^{Stein}$ & 1.0477 & 0.8971 & 0.7219 & 0.4131 & 0.4869 & 0.5688 & 0.1128 & 0.1116 & 0.1013 \\ \hline
$\Xi^{SymStein}$ & 1.0574 & 0.9009 & 0.7213 & 0.4110 & 0.4881 & 0.5680 & 0.1113 & 0.1105 & 0.1010 \\ \hline
$\Xi^{YCM}$ & 0.9867 & 0.8647 & 0.7279 & 0.$\mathbf{4070}$ & $\mathbf{0.4807}$ & 0.5713 & 0.1119 & 0.1106 & 0.1025 \\ \hline
$\Xi^{2S(LP)}$ & 1.5540 & 0.9119 & 0.7247 & 0.4620 & 0.4911 & 0.5758 & $\mathbf{0.1006}$ & $\mathbf{0.1006}$ & $\mathbf{0.1004}$ \\ \hline
$\Xi^{2S(Stein)}$ & 1.5399 & 0.9068 & 0.7264 & 0.4641 & 0.4890 & 0.5778 & 0.1010 & 0.1010 & 0.1008 \\ \hline
$\Xi^{2S(SymStein)}$ & 1.5451 & 0.9092 & 0.7254 & 0.4628 & 0.4898 & 0.5768 & 0.1007 & 0.1007 & 0.1006 \\ \hline
$\Xi^{2S(YCM)}$ & 1.4504 & 0.8861 & 0.7336 & 0.4653 & 0.4848 & 0.5815 & 0.1022 & 0.1022 & 0.1018 \\ \hline
\end{tabular}
\label{table3}
\end{table}

Finally, table~\ref{table4} shows the results for $\mathcal{R}^2_{out}$. Again, for this metric, the results are heterogeneous. For model 1 and MVP strategy, the best performances are obtained by the non-linear estimators, with a tie for $\Xi^{Stein}$ and $\Xi^{SymStein}$.
For the same model 1, it can be seen that the estimator $\Xi^{naive}$ has the best performance in terms of the MVP+ and HRP strategies. Therefore, for this combination of model and strategies, it is sufficient to use the sample estimator to reduce the realized risk.
In model 2, the estimator $\Xi^{LP}$ reduces the $\mathcal{R}^2_{out}$ for both the MVP and HRP strategies. While $\Xi^{YCM}$ does so for MVP+. In the case of model 3, the hierarchical estimator $\Xi^{ALCA}$ has the best performance under the three investment strategies.
\begin{table}[htbp]
\tiny
\centering
\caption{Average $\mathcal{R}^2_{out}$ over $m=1000$ realizations of model 1, model 2, and model 3 on the allocation strategies MVP, MPV+, and HRP. The random matrix samples have dimensions $p=100, n=200$. }
\begin{tabular}{lrrrrrrrrr}
\hline
 & \multicolumn{ 3}{l|}{Model 1} & \multicolumn{ 3}{l|}{Model 2} & \multicolumn{ 3}{l|}{Model 3} \\ \hline
estimator & \multicolumn{1}{l|}{MVP} & \multicolumn{1}{l|}{MVP+} & \multicolumn{1}{l|}{HRP} & \multicolumn{1}{l|}{MVP} & \multicolumn{1}{l|}{MVP+} & \multicolumn{1}{l|}{HRP} & \multicolumn{1}{l|}{MVP} & \multicolumn{1}{l|}{MVP+} & \multicolumn{1}{l|}{HRP} \\ \hline
$\Sigma$ & 0.0100 & 0.0100 & 0.0383 & 0.0047 & 0.0106 & 0.0193 & 0.0268 & 0.0268 & 0.0268 \\ \hline
$\Xi^{naive}$ & 0.0200 & $\mathbf{0.0101}$ & $\mathbf{0.0379}$ & 0.0111 & 0.0121 & 0.0195 & 0.0537 & 0.0420 & 0.0281 \\ \hline
$\Xi^{linear}$ & 0.0134 & 0.0113 & 0.0481 & 0.0079 & 0.0120 & 0.0195 & 0.0404 & 0.0403 & 0.0345 \\ \hline
$\Xi^{ALCA}$ & 0.0136 & 0.0105 & 0.0392 & 0.0093 & 0.0121 & 0.0205 & $\mathbf{0.0277}$ & $\mathbf{0.0277}$ & $\mathbf{0.0274}$ \\ \hline
$\Xi^{LP}$ & 0.0134 & 0.0103 & 0.0402 & $\mathbf{0.0057}$ & 0.0117 & $\mathbf{0.0193}$ & 0.0381 & 0.0379 & 0.0318 \\ \hline
$\Xi^{Stein}$ & $\mathbf{0.0133}$ & 0.0104 & 0.0413 & 0.0060 & 0.0117 & 0.0195 & 0.0385 & 0.0379 & 0.0305 \\ \hline
$\Xi^{SymStein}$ & $\mathbf{0.0133}$ & 0.0104 & 0.0407 & 0.0058 & 0.0117 & 0.0194 & 0.0383 & 0.0378 & 0.0311 \\ \hline
$\Xi^{YCM}$ & 0.0135 & 0.0111 & 0.0466 & 0.0059 & $\mathbf{0.0114}$ & 0.0198 & 0.0391 & 0.0385 & 0.0302 \\ \hline
$\Xi^{2S(LP)}$ & 0.0143 & 0.0109 & 0.0420 & 0.0097 & 0.0119 & 0.0203 & 0.0317 & 0.0317 & 0.0317 \\ \hline
$\Xi^{2S(Stein)}$ & 0.0146 & 0.0110 & 0.0432 & 0.0100 & 0.0119 & 0.0205 & 0.0305 & 0.0305 & 0.0304 \\ \hline
$\Xi^{2S(SymStein)}$ & 0.0144 & 0.0109 & 0.0425 & 0.0098 & 0.0119 & 0.0204 & 0.0311 & 0.0311 & 0.0310 \\ \hline
$\Xi^{2S(YCM)}$ & 0.0157 & 0.0118 & 0.0492 & 0.0103 & 0.0118 & 0.0209 & 0.0303 & 0.0303 & 0.0304 \\ \hline
\end{tabular}
\label{table4}
\end{table}

\section{Empirical data}

In this section, we consider empirical data to analyze the behavior of the covariance estimators and investment strategies in a real scenario.
We considered the adjusted close prices of the assets that comprise the S\&P500 for the elapsed period from 2012-01-03 to 2021-12-30 at a daily frequency.
The source of data is the free repository of Yahoo Finance.
We only included assets having less than 5\% missing values and applied the standard imputation strategy in empirical finance of repeating the last non-null value~\footnote{We have used the same dataset and preprocessing procedure presented in~\cite{garcia2024random}}.
We end up with a set of $p=441$ return times series with a length of $T=2515$ transaction days, where returns were computed using Eq.~\ref{returns}.

\begin{figure}[hbtp]
    \centering
    \begin{subfigure}[b]{0.45\textwidth}
        \includegraphics[scale=0.25]{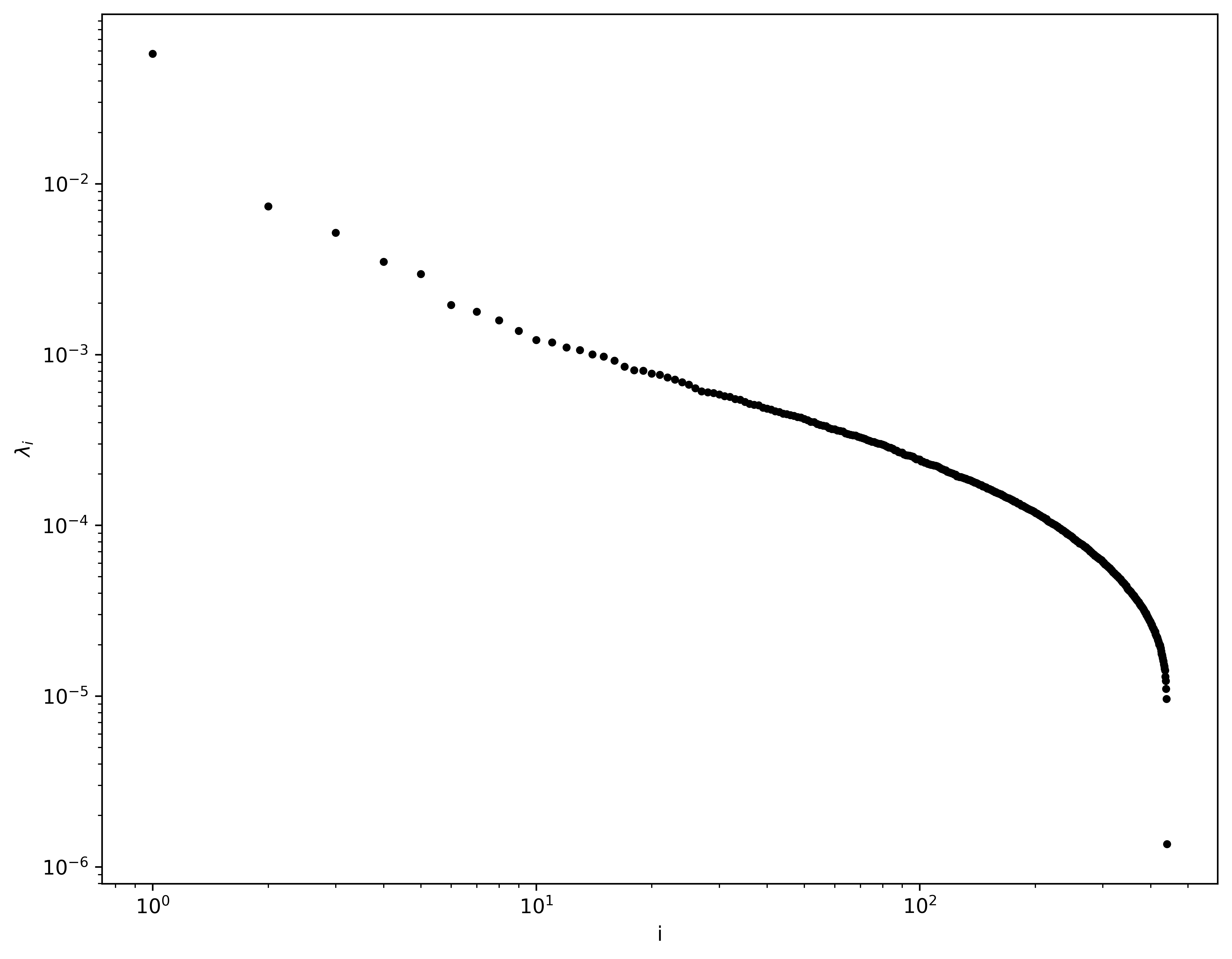}
        \caption{}
    \end{subfigure}
    \begin{subfigure}[b]{0.45\textwidth}
        \includegraphics[scale=0.25]{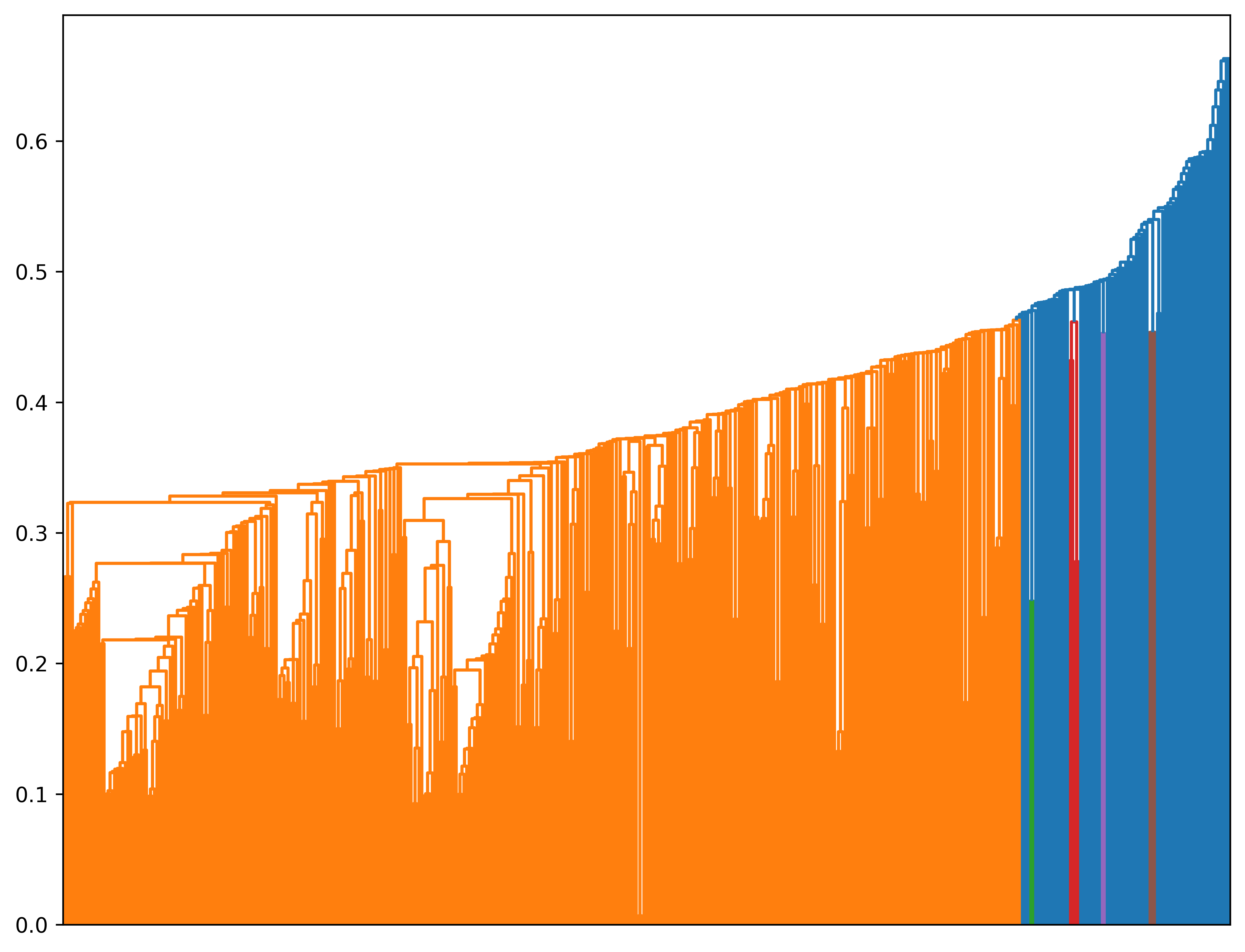}
        \caption{}
    \end{subfigure}
    \caption{Empirical data. (a) Scree plot of ordered eigenvalues. (b) Dendrogram under the same methodology of the hierarchical estimator with single linkage (SLCA).}
    \label{fig5}
\end{figure}

Fig.~\ref{fig5} shows the scree plot of ordered eigenvalues(a) and dendrogram (b) of the empirical covariance matrix. Here, the dendrogram is computed using the same methodology as the hierarchical estimator but integrating SLCA instead of ALCA.
We can see that the scree plot resembles model 1 and that the dendrogram presents a high degree of nested hierarchies.

To analyze the temporal behavior of performance metrics on the empirical data we have generated moving windows.
We build subsamples of size $n=2p$ shifted by steps of the size of $\Delta t=10$, obtaining a total of $m=75$ data windows.
Then, we applied the covariance estimators discussed in section~\ref{estimators} to each data window and computed the optimized weights under the different investment strategies. 
Here, $\mathcal{L}$ and $\mathcal{HHI}$ are obtained in-sample by construction. While $\mathcal{RDI}$ and $\mathcal{R}^2_{out}$ are calculated out-sample using the subsequent data window. 

\begin{figure}[hbtp]
    \centering
        \includegraphics[scale=0.5]{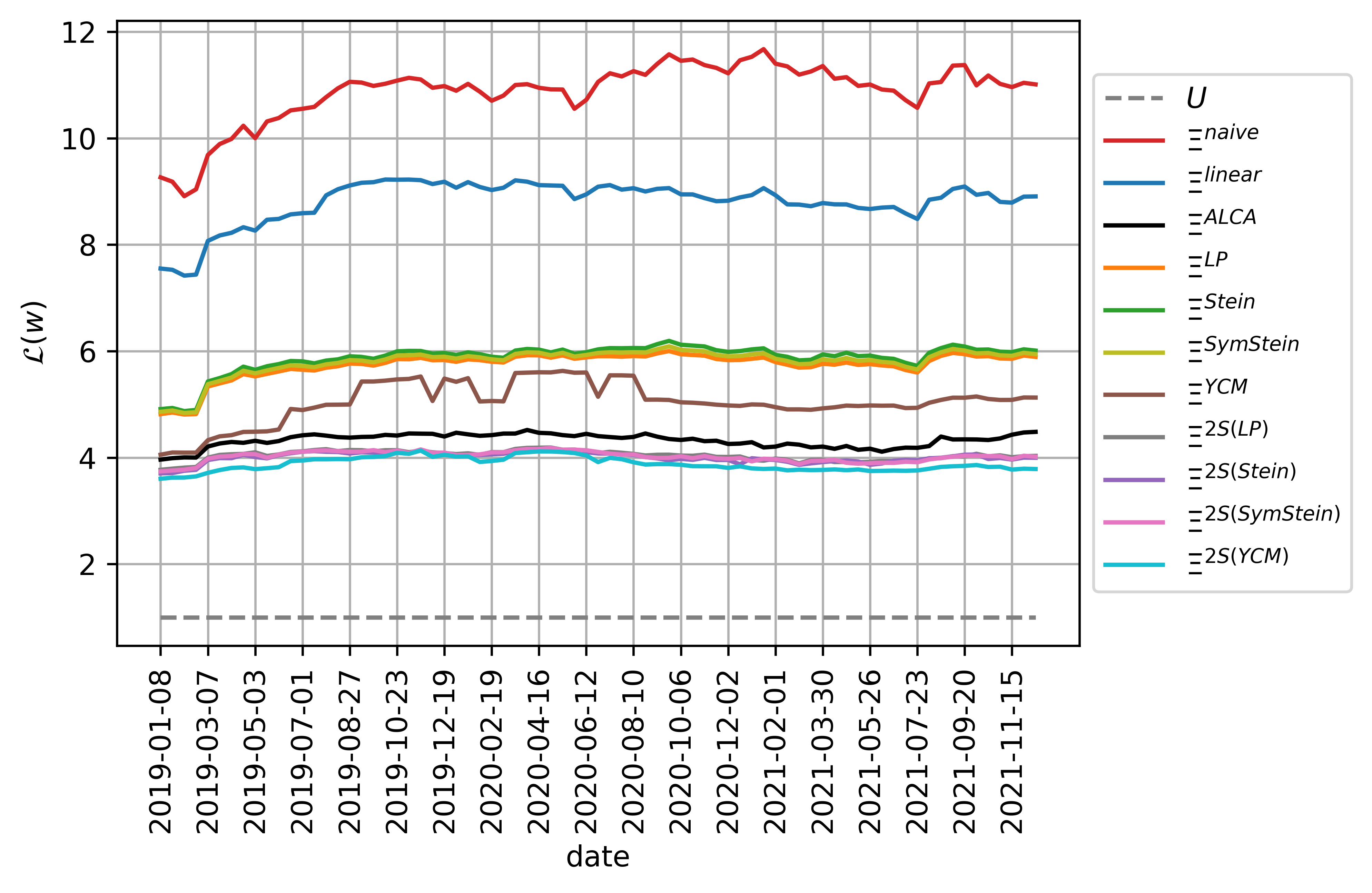}
    \caption{Performance $\mathcal{L}$  on the empirical data for the investment strategy MV. The moving windows are composed of $p=441$ assets and $n=882$ transaction days. The x-axis denotes the end date of the out-sample window.}
    \label{fig6}
\end{figure}

In Fig~\ref{fig6} we can see the behavior of $\mathcal{L}$ in the empirical data, where it is observed that the leverage reaches its lowest values with the estimator $\Xi^{2S(YCM)}$. For this metric, the performance of the MV+ and HRP strategies was not plotted since leverage is not allowed by construction, i.e., $\mathcal{L}=1$ for all time windows.

\begin{figure}[hbtp]
    \centering
    \begin{subfigure}[b]{0.45\textwidth}
        \includegraphics[scale=0.5]{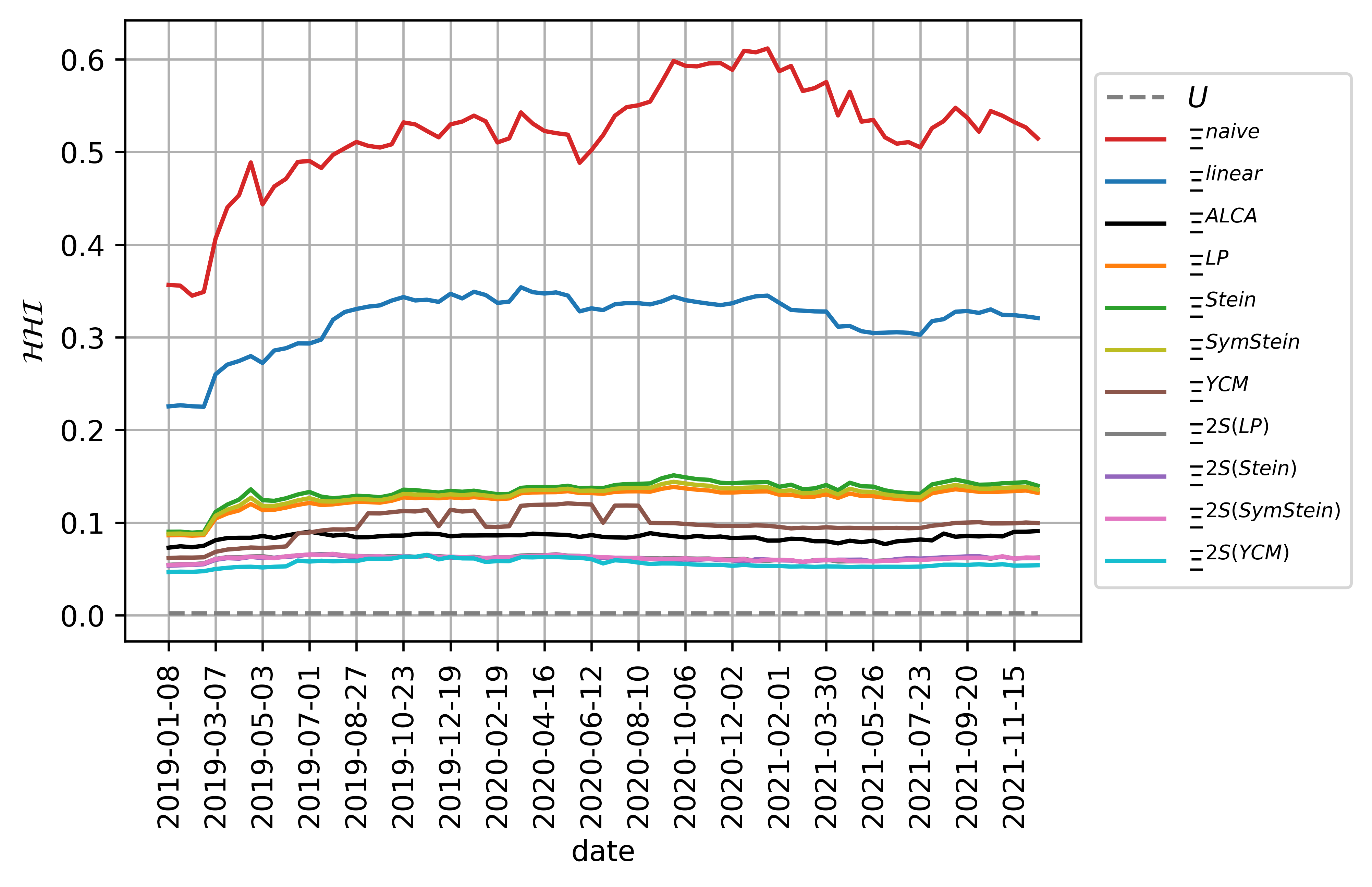}
        \caption{}
    \end{subfigure}\\
    \begin{subfigure}[b]{0.45\textwidth}
        \includegraphics[scale=0.5]{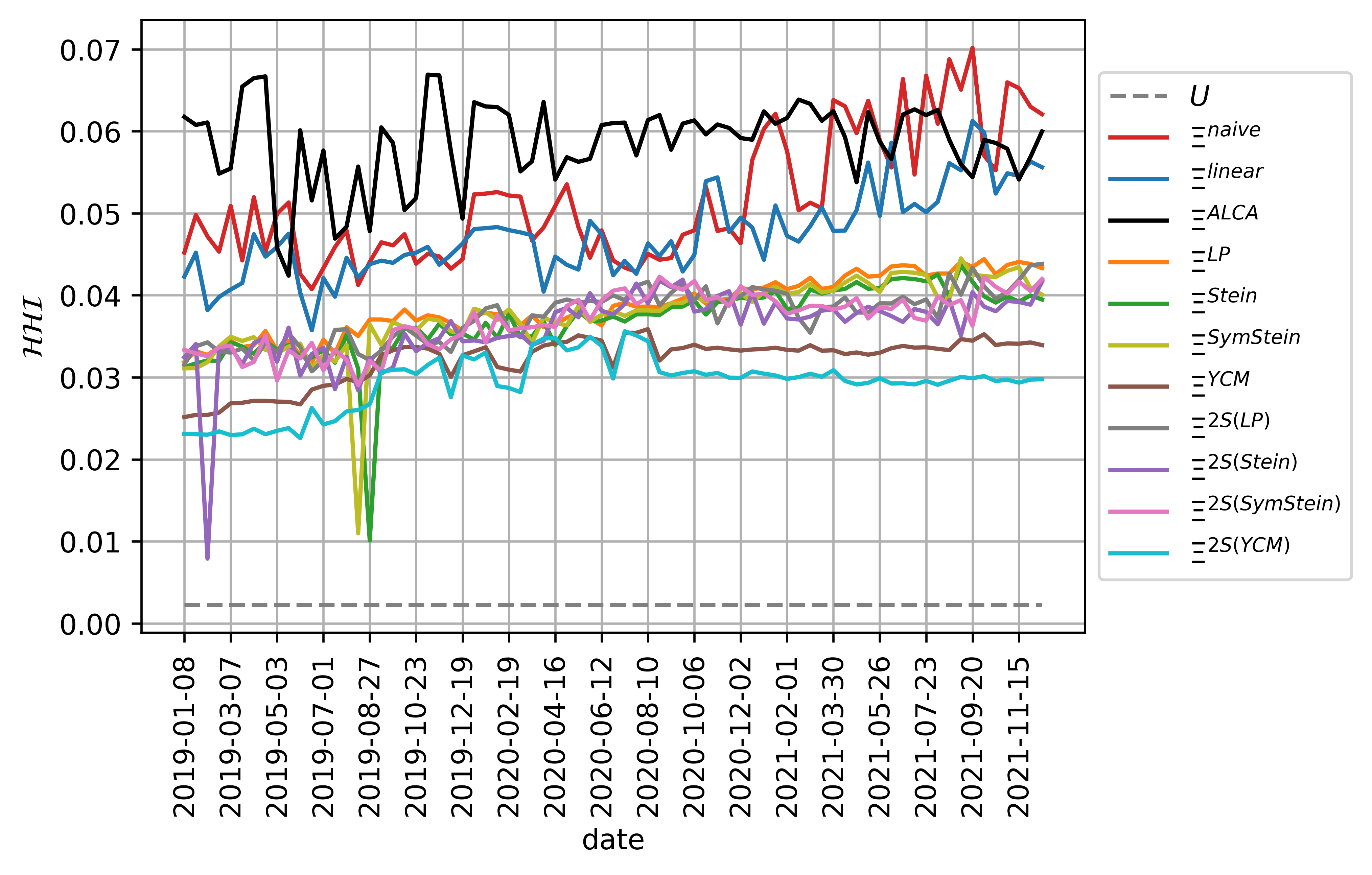}
        \caption{}
    \end{subfigure}\\
    \begin{subfigure}[b]{0.45\textwidth}
        \includegraphics[scale=0.5]{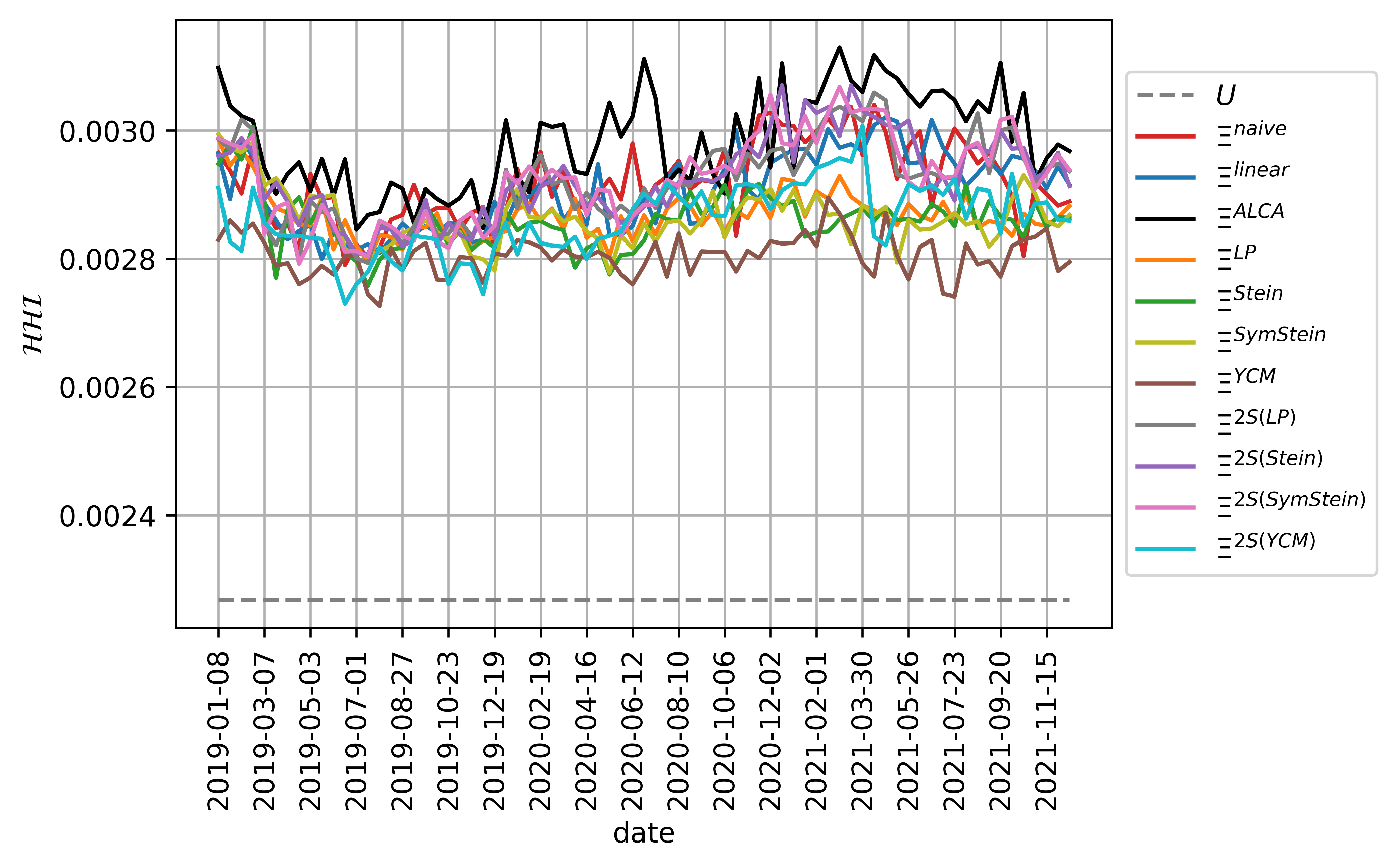}
        \caption{}
    \end{subfigure}
    \caption{Performance $\mathcal{HHI}$   on the empirical data for the investment strategy (a)MV, (b)MV+, and (c)HRP. The moving windows are composed of $p=441$ assets and $n=882$ transaction days. The x-axis denotes the end date of the out-sample window.}
    \label{fig7}
\end{figure}

In Fig~\ref{fig7} the behavior of the metric $\mathcal{HHI}$ can be visualized. Again, the estimator $\Xi^{2S(YCM)}$ minimizes the value of $\mathcal{HHI}$ for the MV strategy consistently, and most of the time, and to a greater extent for MV+. However, for the HRP strategy, the behavior is somewhat heterogeneous over time, but the estimators $\Xi^{2S(YCM)}$ and $\Xi^{YCM}$ stand out as the ones with the best performances.

\begin{figure}[hbtp]
    \centering
    \begin{subfigure}[b]{0.45\textwidth}
        \includegraphics[scale=0.5]{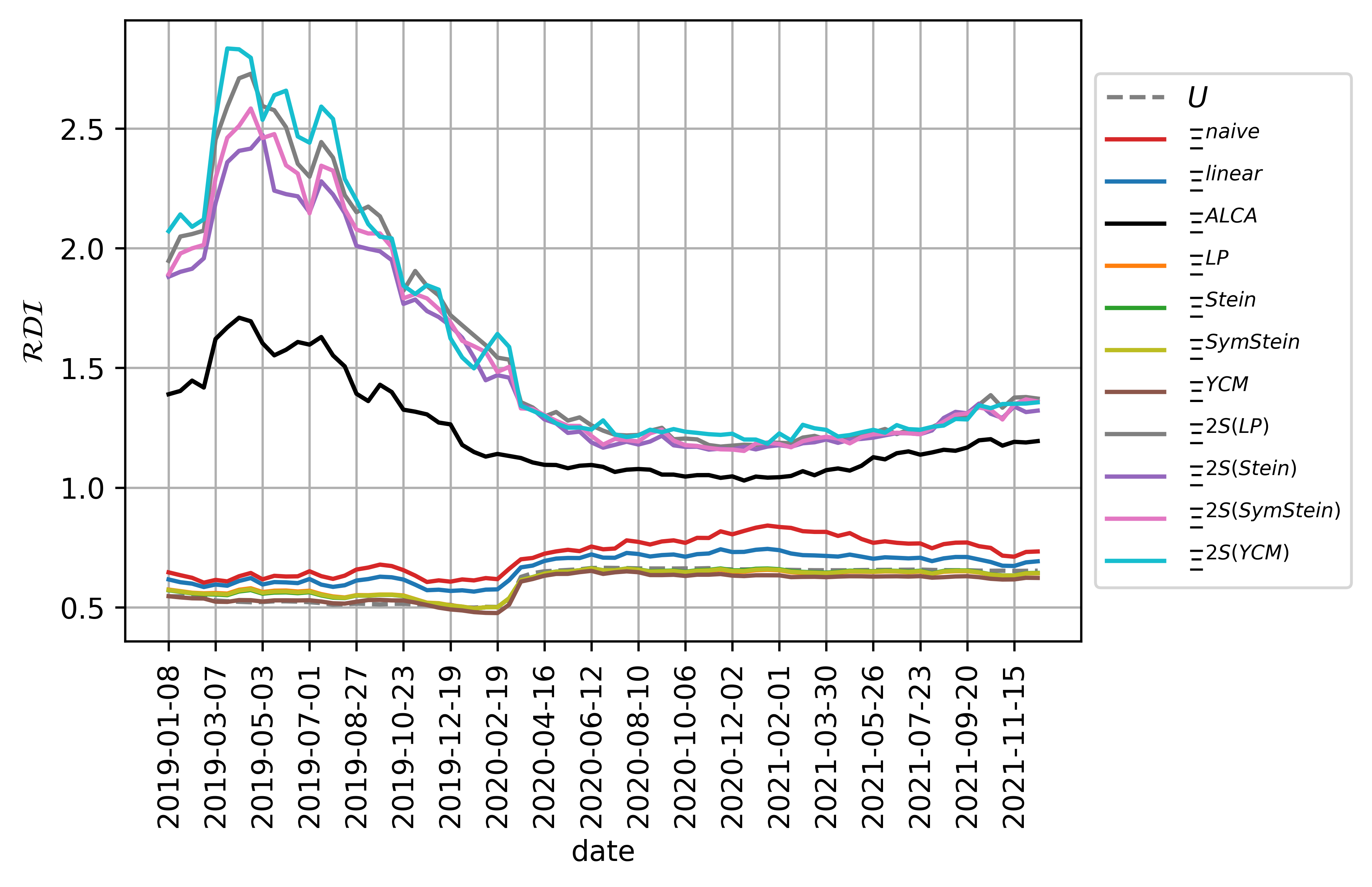}
        \caption{}
    \end{subfigure}\\
    \begin{subfigure}[b]{0.45\textwidth}
        \includegraphics[scale=0.5]{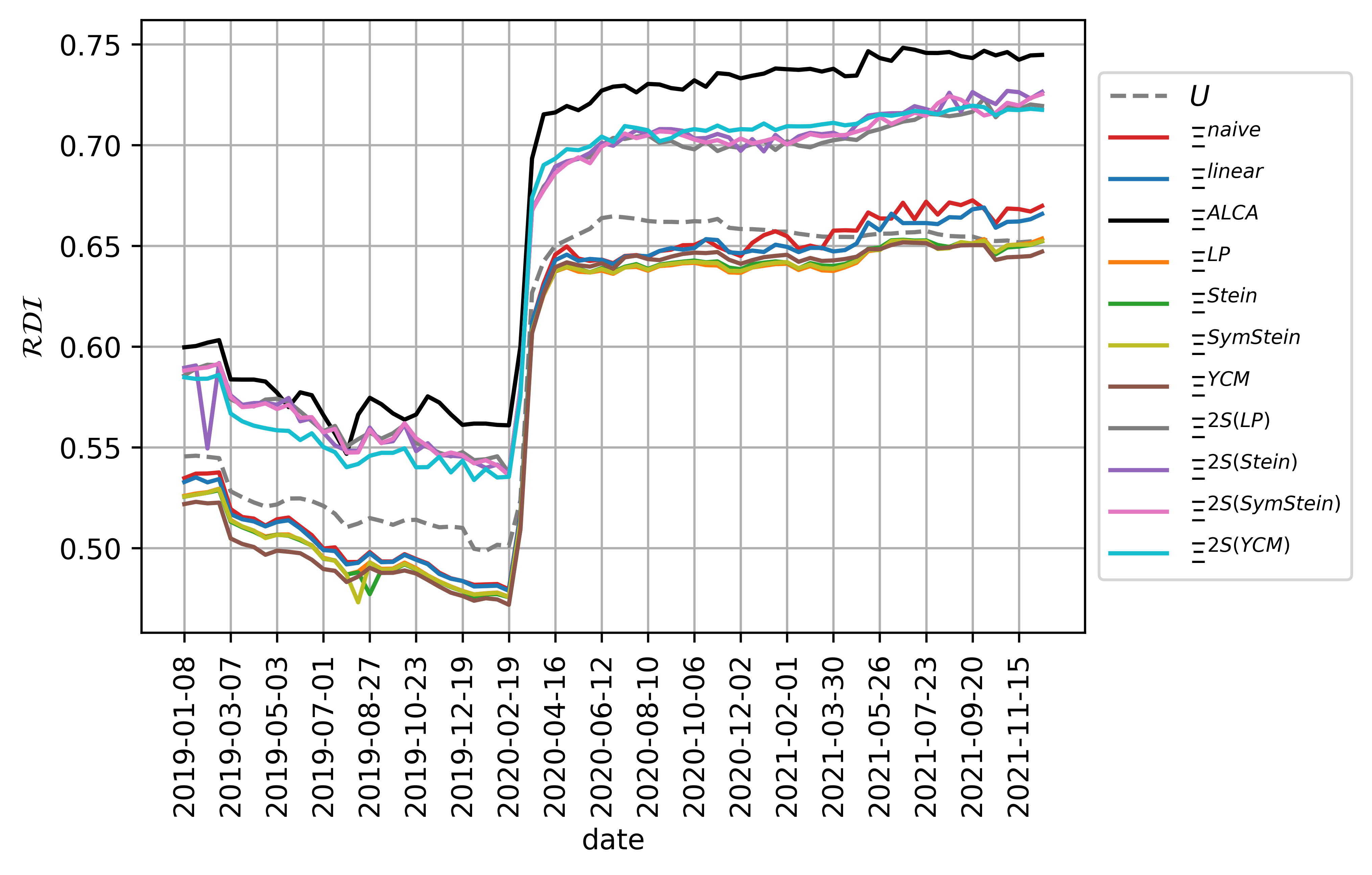}
        \caption{}
    \end{subfigure}\\
    \begin{subfigure}[b]{0.45\textwidth}
        \includegraphics[scale=0.5]{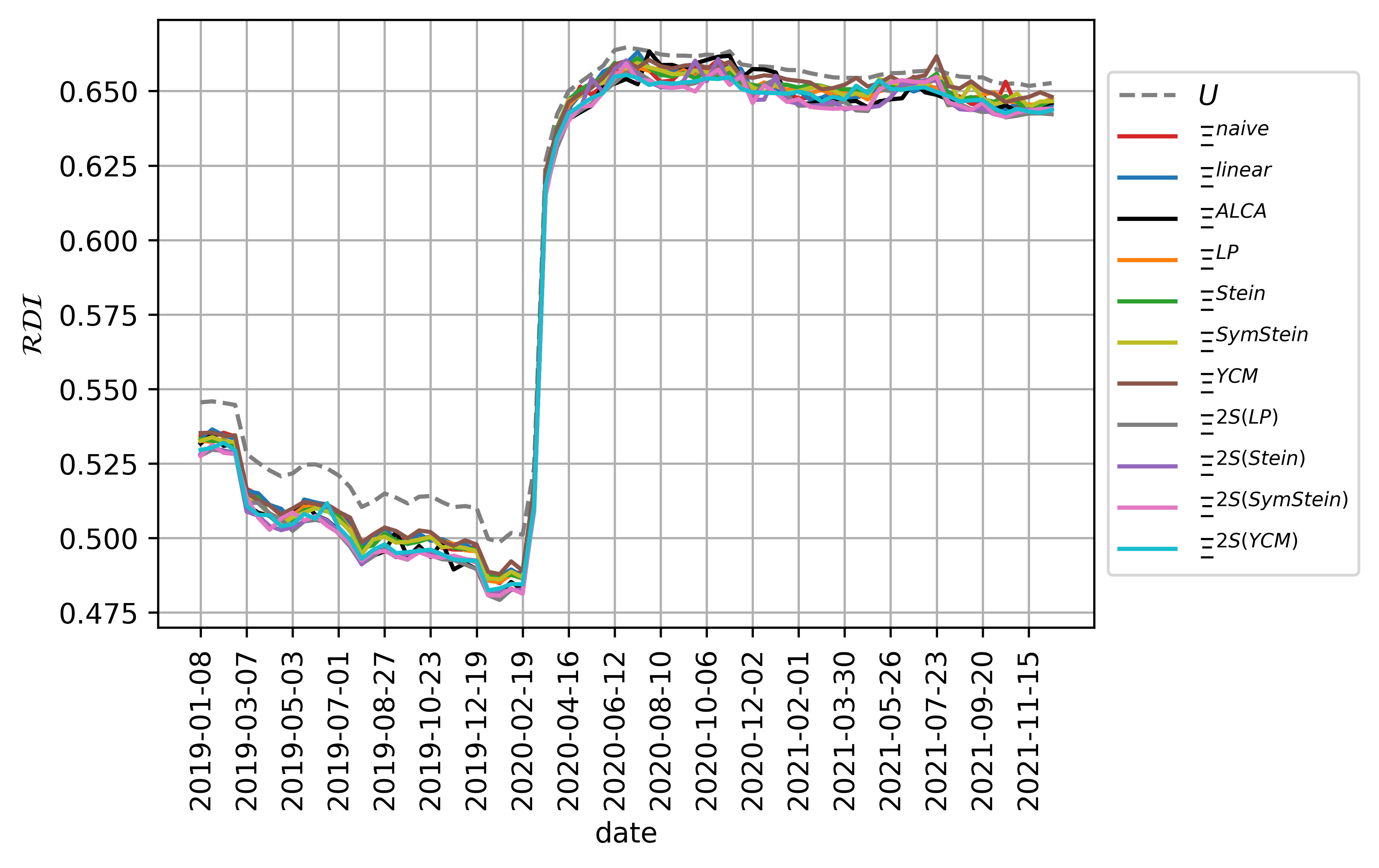}
        \caption{}
    \end{subfigure}\\
    \caption{Performance $\mathcal{RDI}$  on the empirical data for the investment strategy (a)MV, (b)MV+, and (c)HRP. The moving windows are composed of $p=441$ assets and $n=882$ transaction days. The x-axis denotes the end date of the out-sample window.}
    \label{fig8}
\end{figure}

Fig.\ref{fig8} shows the behavior of the metric $\mathcal{RDI}$ where it can be seen that for the MV strategy, the best performance is consistently taken by the estimator $\Xi^{YCM}$, although the non-linear estimators have a very similar performance. Interestingly, for the MV+ strategy, the best performance is associated with the estimator $\Xi^{YCM}$ until around March 2020, and then the non-linear estimators have a slight advantage after that. In the case of the HRP strategy, there is a similar inflection point, but in this case, the competing estimators are the two-step ones. It is important to remember that March 2020 was marked by the event of the beginning of the COVID-19 pandemic so it is natural that $\mathcal{RDI}$ increases after this critical point.

\begin{figure}[hbtp]
    \centering
    \begin{subfigure}[b]{0.45\textwidth}
        \includegraphics[scale=0.5]{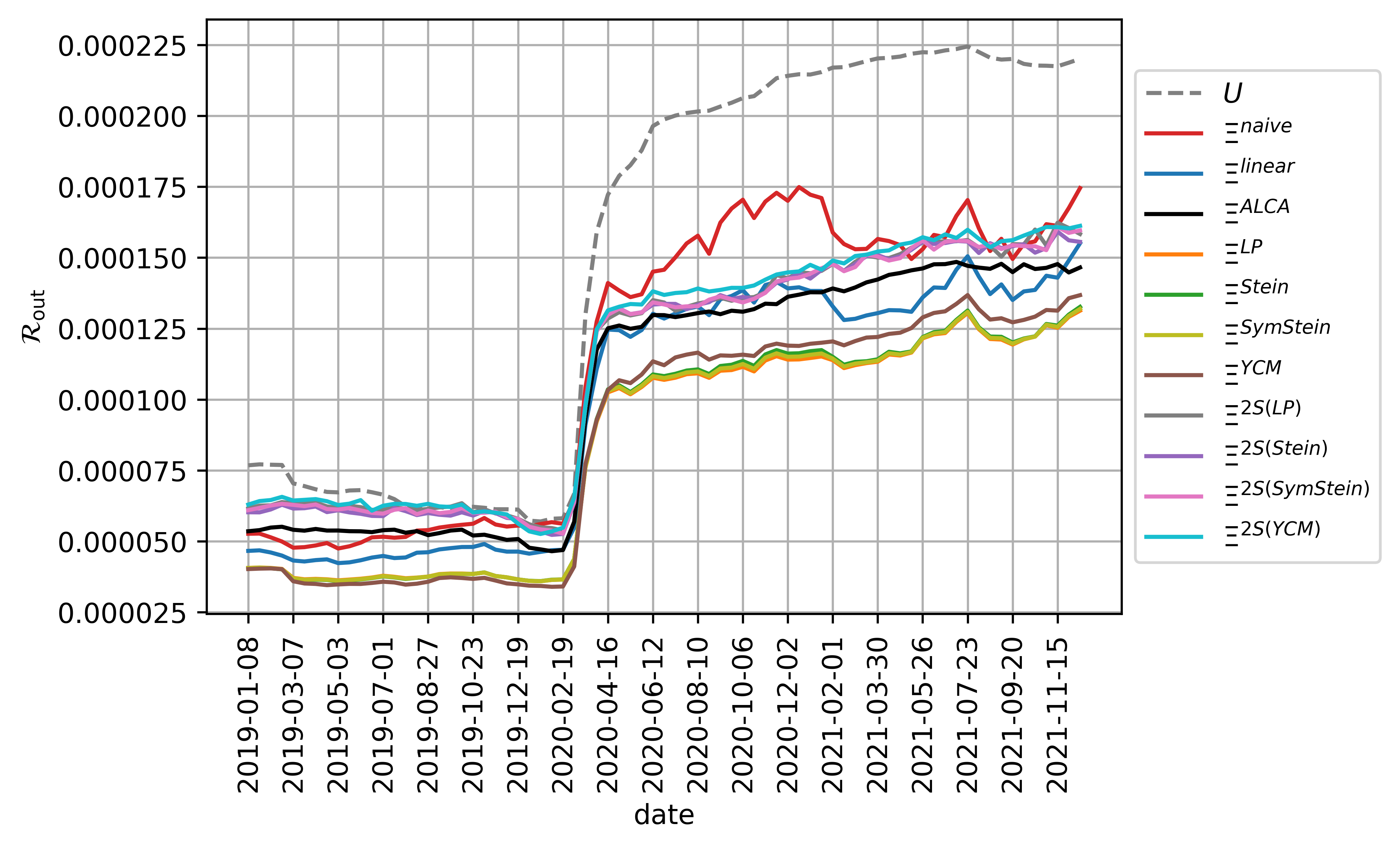}
        \caption{}
    \end{subfigure}\\
    \begin{subfigure}[b]{0.45\textwidth}
        \includegraphics[scale=0.5]{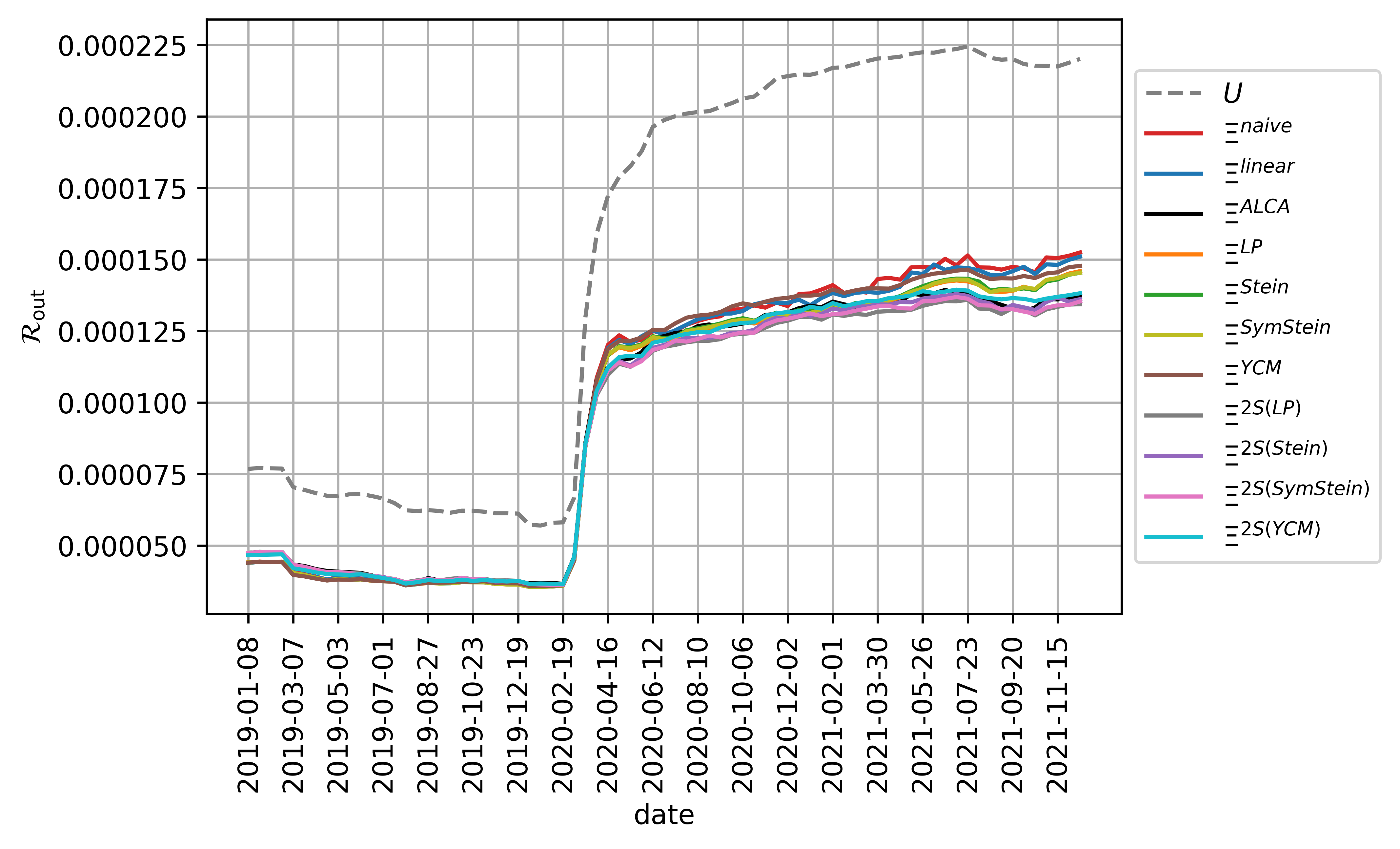}
        \caption{}
    \end{subfigure}\\
    \begin{subfigure}[b]{0.45\textwidth}
        \includegraphics[scale=0.5]{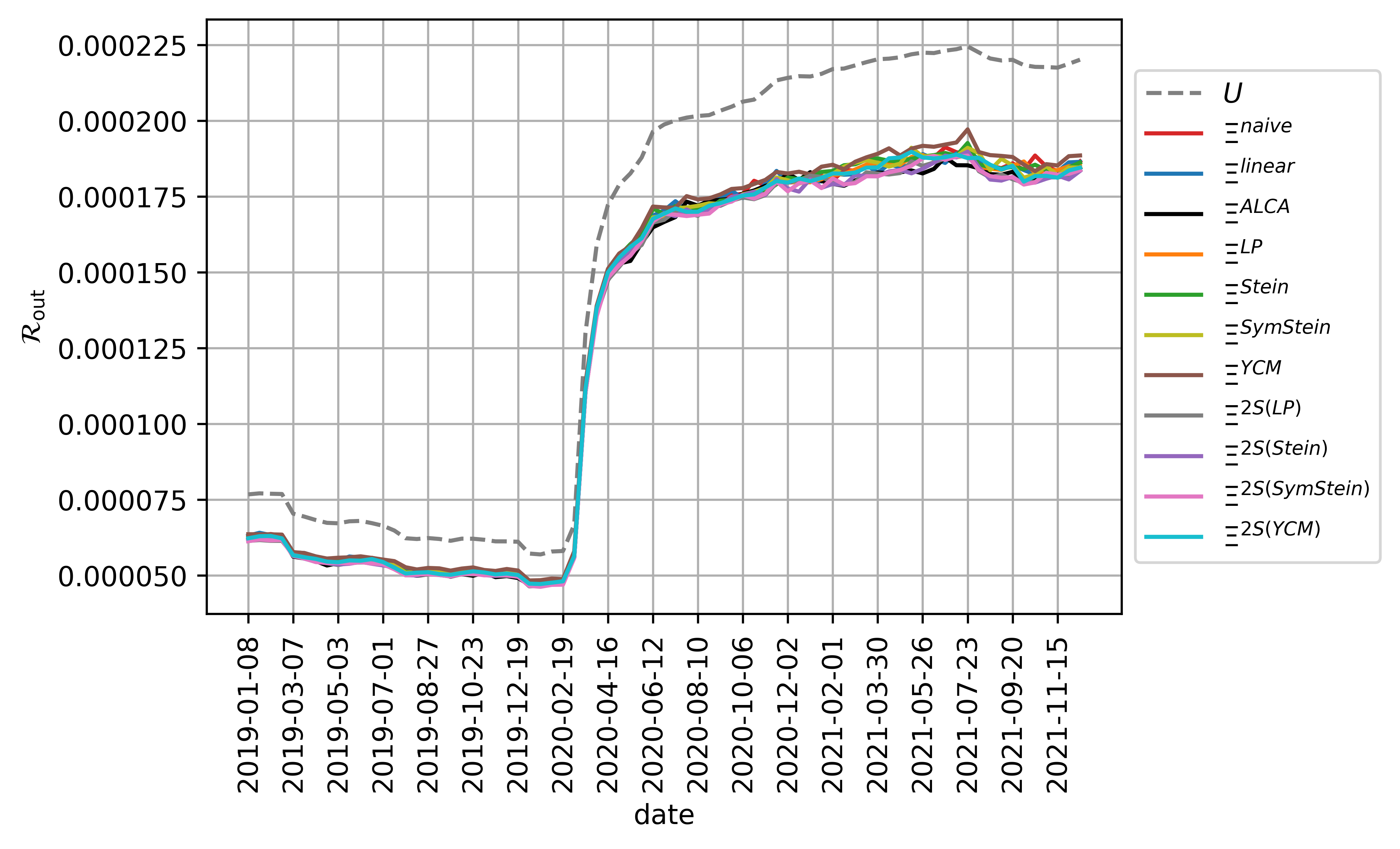}
        \caption{}
    \end{subfigure}
    \caption{Performance $\mathcal{R}_{out}^2$   on the empirical data for the investment strategy (a)MV, (b)MV+, and (c)HRP. The moving windows are composed of $p=441$ assets and $n=882$ transaction days. The x-axis denotes the end date of the out-sample window.}
    \label{fig9}
\end{figure}

Finally, Fig~\ref{fig9} shows the behavior of $\mathcal{R}_{out}^2$. The most notable thing about these graphs is that all estimators improve the realized risk compared to the uniform portfolio strategy. In the case of the MV strategy, the $\Xi^{YCM}$ estimator can be seen to have the best performance before the start of the pandemic, while again the non-linear estimators have a slight advantage after this event. In the case of the MV+ strategy, there is no clear winner before March 2020. However, after this turbulence, it can be noted that the two-step estimators have a better performance. In the case of HRP, the results are even more heterogeneous and fluctuating with a barely highlighting the two-step estimator over the others. The clear pattern in the three strategies is that after the start of the pandemic, the effect of the estimators is less bounded, and spread after that, managing to differentiate one from the other.

\begin{figure}[hbtp]
    \centering
    \begin{subfigure}[b]{0.4\textwidth}
        \includegraphics[scale=0.25]{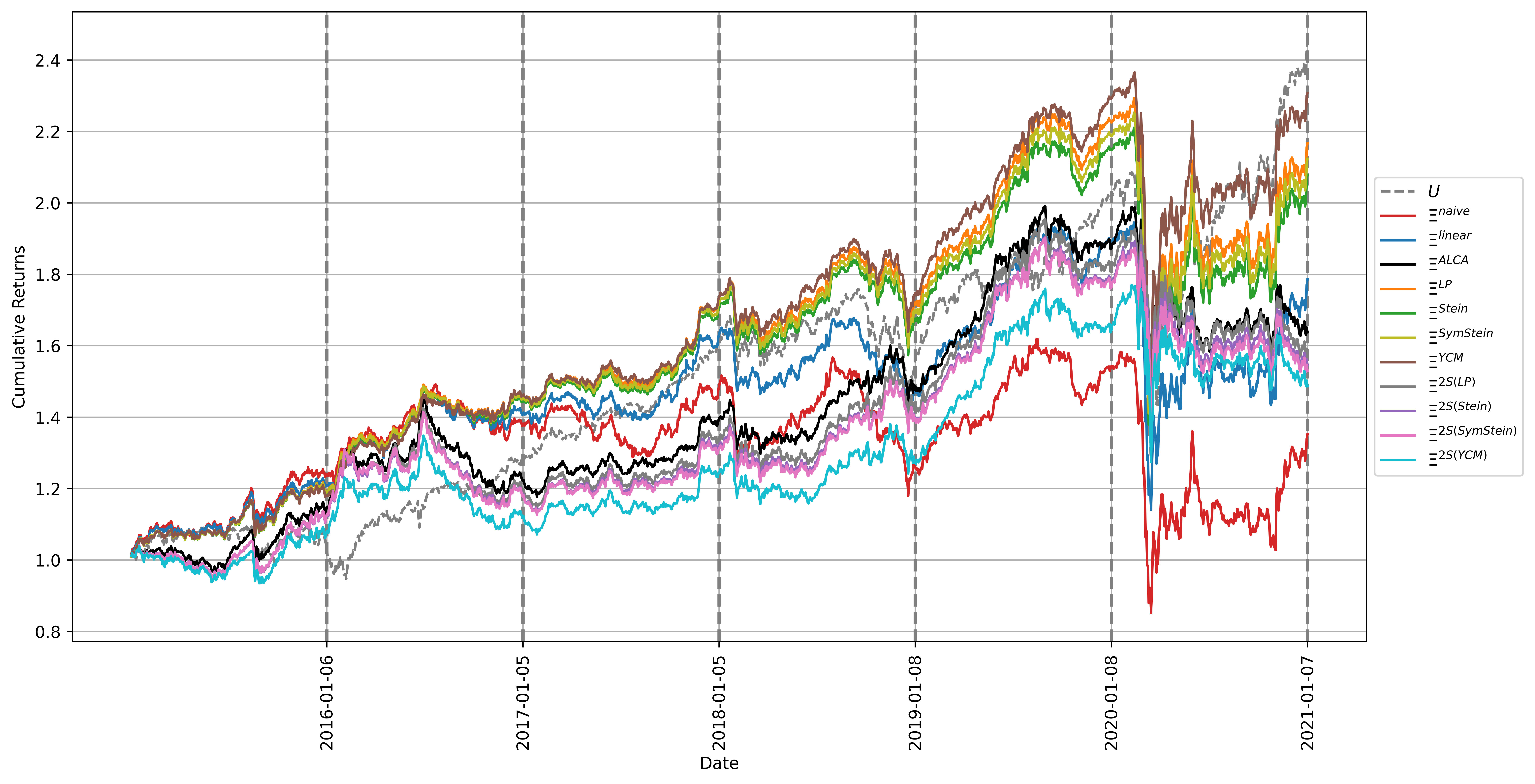}
        \caption{}
    \end{subfigure}\\
    \begin{subfigure}[b]{0.4\textwidth}
        \includegraphics[scale=0.25]{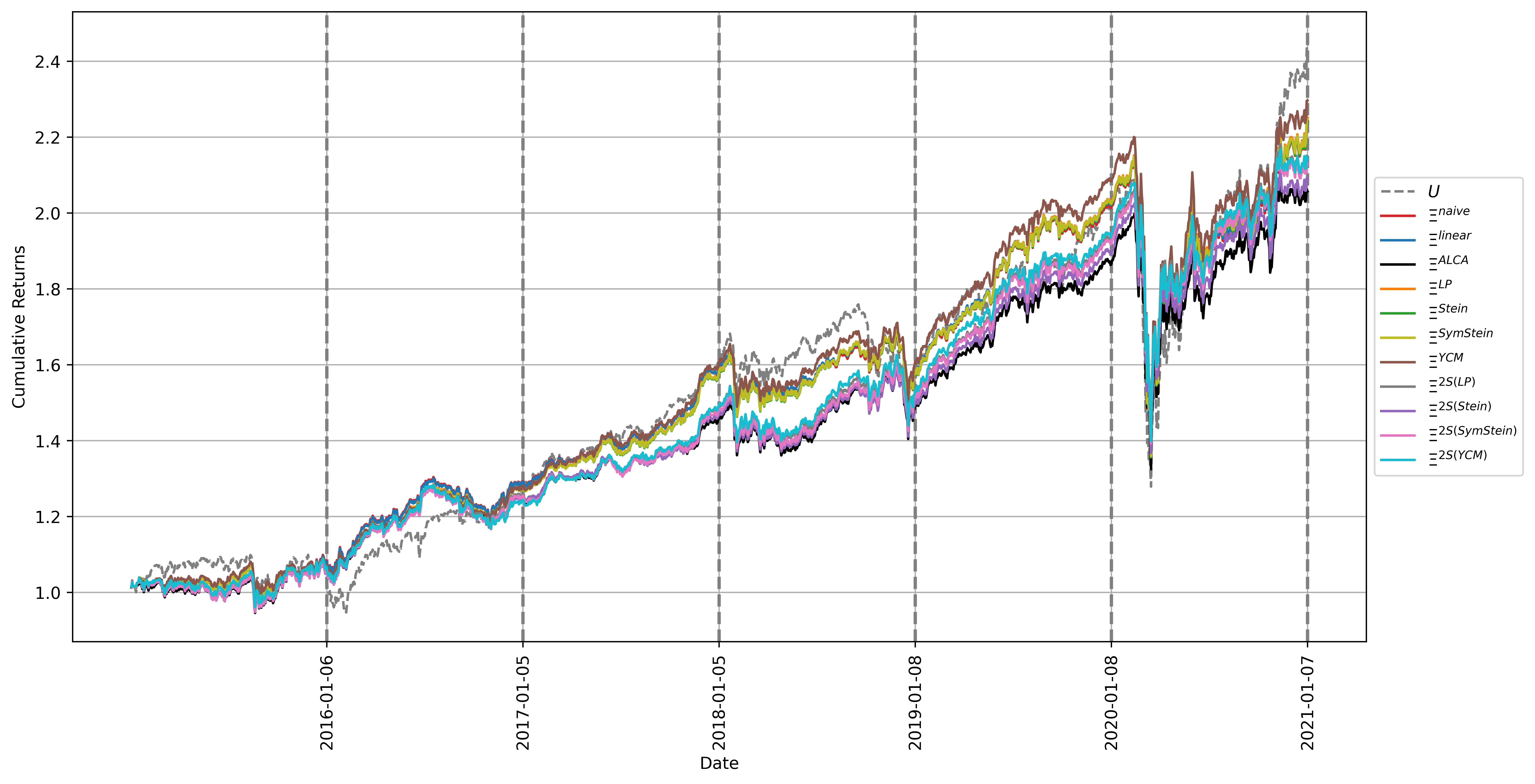}
        \caption{}
    \end{subfigure}\\
    \begin{subfigure}[b]{0.4\textwidth}
        \includegraphics[scale=0.25]{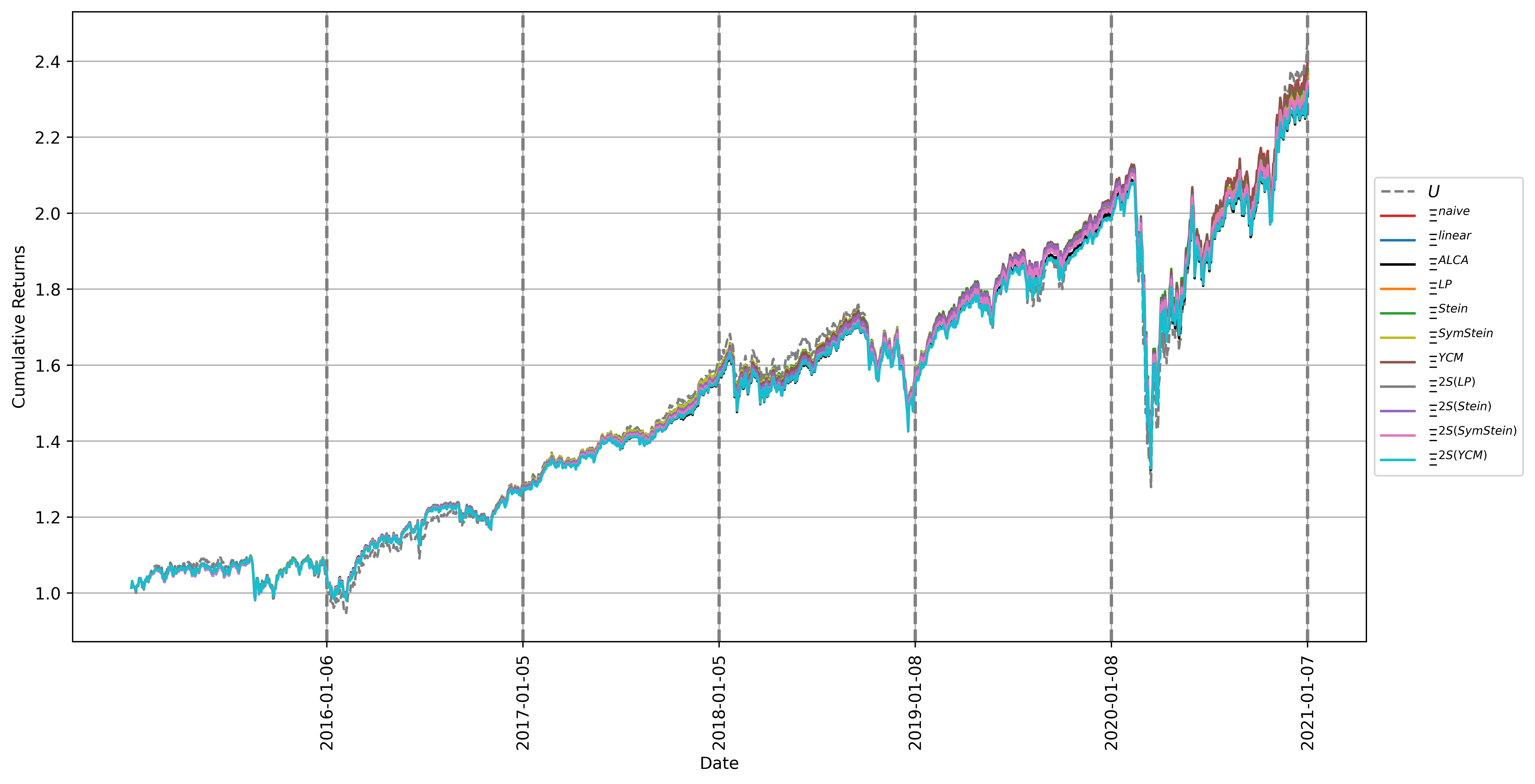}
        \caption{}
    \end{subfigure}
    \caption{Walk-forward cumulative returns on empirical data for (a) MVP, (b) MVP+, and (c) HRP. The weights are optimized with $T_{in}=756$ days, applied over $T_{out}=252$, and rebalancing every $\Delta T = 252$ days.}
    \label{fig10}
\end{figure}

As a practical example, a walk-forward analysis (backtesting) was performed using the equivalent of three years of historical returns to estimate the covariance matrix and optimize the allocation weights. The investment strategy was applied out-sample for one year and rebalanced yearly. In total, we rebalance the portfolio six times between 2015 and 2021 around the start of each year.
Fig.\ref{fig10} shows the cumulative return combination of estimator and investment strategy, where each year was considered composed of 252 transaction days, and the rebalancing dates have been delimited by grey dashed vertical lines. It is observed that before the pandemic declaration, the MVP strategy obtained the best cumulative performance with the estimator $\Xi^{YCM}$, closely following the non-linear estimators. Nevertheless, at the end of the period, the most lucrative strategy turns out to be the uniform portfolio. In the case of the MVP+ strategy, there is less difference in the behavior of the cumulative performance over time. In some periods the uniform portfolio turns out to be the best and in others, the performance is favored by the estimator $\Xi^{YCM}$. But finally, at the end of the investment, the uniform portfolio wins. It is remarkable to observe that in the case of the HRP strategy, the cumulative returns under the different estimators differ very little. However, it is notable that its performance is very close to that of the uniform portfolio, suggesting that the HRP strategy can be more profitable than the previous ones independently of the estimators.

Table \ref{table5} shows the performance metrics of walk-forward analysis on the empirical data for the investment strategies. The annual return and annual volatility are given by the average and standard deviation forward daily return over the analyzed period multiplied by factor 252 to obtain a proxy of yearly quantities. The Sharpe ratio is simply the ratio between the annual return and the annual volatility. The Maximum (Max) Drawdown measures the largest peak-to-trough decline in the cumulative return of the investment portfolio expressed as a percentage. This measure represents the maximum loss presented in the investment. The Sortino ratio is a variant measure of the Sharpe ratio, in which only the standard deviation of the negative returns is taken into account. It gives us an idea of the downside risks that accompany a stock. Finally, turnover measures the absolute difference between the asset weights at consecutive periods normalized by the number of periods. A high value is associated with high transaction fees.

In terms of Annual Return the Uniform portfolio obtained the highest average value over the the period. Nevertheless, the risk associated with this strategy is also high. The annual volatility is reduced by all the estimators. The minimum volatility is reached by the $\Xi^{2S(LP)}$ estimator under the MVP+ strategy. The highest Sharpe ratio is given by $\Xi^{YCM}$ for MVP strategy, yet it is less than 1, which is the baseline bound to consider a good investment. Moreover, the Maximum Drawdown is reduced to -27.93\% by  $\Xi^{2S(YCM)}$ under the MVP strategy. Further, the Sortino ratio reaches the value of 1.04 again for $\Xi^{YCM}$ under MVP. Then, in situations of risk aversion, it is the best option. Further, the turnover is minimized by the $\Xi^{2S(YCM)}$ for the HRP strategy. These portfolios have very smooth changes compared to the naive estimator for example. Hence the transaction fees are also reduced. 

\begin{table}[htbp]
\caption{Performance metrics of walk-forward analysis on the empirical data for MVP,  MVP+, and HRP investment strategies. The weights are optimized with $T_{in}=756$ days, applied over $T_{out}=252$, and rebalancing every $\Delta T = 252$ days.}
\tiny
\centering
\begin{tabular}{lllllll}
\hline
Estimator & Annual Return & Annual Volatility & Sharpe Ratio & Max Drawdown & Sortino Ratio & Turnover \\
\hline
$U$ & $\mathbf{16.96\%}$ & 19.86\% & 0.85 & -38.73\% & 0.97 & 0.00 \\ \hline
$\Xi^{naive}$(MVP) & 7.21\% & 20.93\% & 0.34 & -47.40\% & 0.43 & 11.78 \\ \hline
$\Xi^{linear}$(MVP) & 11.34\% & 18.28\% & 0.62 & -41.34\% & 0.74 & 7.37 \\ \hline
$\Xi^{ALCA}$(MVP) & 9.66\% & 16.94\% & 0.57 & -30.39\% & 0.70 & 2.10 \\ \hline
$\Xi^{LP}$(MVP) & 14.35\% & 17.05\% & 0.84 & -38.98\% & 0.94 & 3.94 \\ \hline
$\Xi^{Stein}$(MVP) & 13.67\% & 17.21\% & 0.79 & -39.53\% & 0.89 & 4.28 \\ \hline
$\Xi^{SymStein}$(MVP) & 14.03\% & 17.12\% & 0.82 & -39.24\% & 0.92 & 4.09 \\ \hline
$\Xi^{YCM}$(MVP) & 15.27\% & 16.29\% & $\mathbf{0.94}$ & -36.51\% & $\mathbf{1.04}$ & 3.00 \\ \hline
$\Xi^{2S(LP)}$(MVP) & 9.05\% & 17.26\% & 0.52 & -29.28\% & 0.67 & 1.79 \\ \hline
$\Xi^{2S(Stein)}$(MVP) & 8.75\% & 17.26\% & 0.51 & -29.24\% & 0.64 & 1.74 \\ \hline
$\Xi^{2S(SymStein)}$(MVP) & 8.59\% & 17.35\% & 0.50 & -30.03\% & 0.63 & 1.75 \\ \hline
$\Xi^{2S(YCM)}$(MVP) & 8.08\% & 17.18\% & 0.47 & $\mathbf{-27.93\%}$ & 0.60 & 1.51 \\ \hline
$\Xi^{naive}$(MVP+) & 15.00\% & 17.44\% & 0.86 & -37.45\% & 0.95 & 0.92 \\ \hline
$\Xi^{linear}$(MVP+) & 14.92\% & 17.39\% & 0.86 & -37.35\% & 0.95 & 0.90 \\ \hline
$\Xi^{ALCA}$(MVP+) & 13.39\% & 16.48\% & 0.81 & -33.72\% & 0.93 & 0.77 \\ \hline
$\Xi^{LP}$(MVP+) & 15.01\% & 17.16\% & 0.87 & -36.67\% & 0.96 & 0.83 \\ \hline
$\Xi^{Stein}$(MVP+) & 14.91\% & 17.17\% & 0.87 & -36.79\% & 0.96 & 0.84 \\ \hline
$\Xi^{SymStein}$(MVP+) & 14.95\% & 17.17\% & 0.87 & -36.73\% & 0.96 & 0.83 \\ \hline
$\Xi^{YCM}$(MVP+) & 15.32\% & 17.04\% & 0.90 & -36.67\% & 0.98 & 0.76 \\ \hline
$\Xi^{2S(LP)}$(MVP+) & 14.01\% & $\mathbf{16.11\%}$ & 0.87 & -32.07\% & 1.00 & 0.66 \\ \hline
$\Xi^{2S(Stein)}$(MVP+) & 13.58\% & 16.22\% & 0.84 & -32.52\% & 0.96 & 0.66 \\ \hline
$\Xi^{2S(SymStein)}$(MVP+) & 13.88\% & 16.14\% & 0.86 & -32.18\% & 0.98 & 0.66 \\ \hline
$\Xi^{2S(YCM)}$(MVP+) & 14.04\% & 16.35\% & 0.86 & -32.74\% & 0.97 & 0.60 \\ \hline
$\Xi^{naive}$(HRP) & 16.21\% & 18.15\% & 0.89 & -36.13\% & 1.00 & 0.27 \\ \hline
$\Xi^{linear}$(HRP) & 16.13\% & 18.25\% & 0.88 & -36.34\% & 0.99 & 0.28 \\ \hline
$\Xi^{ALCA}$(HRP) & 15.67\% & 18.27\% & 0.86 & -36.56\% & 0.95 & 0.29 \\ \hline
$\Xi^{LP}$(HRP) & 16.03\% & 18.21\% & 0.88 & -36.51\% & 0.98 & 0.26 \\ \hline
$\Xi^{Stein}$(HRP) & 16.10\% & 18.15\% & 0.89 & -36.09\% & 0.99 & 0.27 \\ \hline
$\Xi^{SymStein}$(HRP) & 15.97\% & 18.17\% & 0.88 & -36.28\% & 0.98 & 0.26 \\ \hline
$\Xi^{YCM}$(HRP) & 16.20\% & 18.32\% & 0.88 & -36.49\% & 0.99 & 0.25 \\ \hline
$\Xi^{2S(LP)}$(HRP) & 15.84\% & 18.16\% & 0.87 & -36.12\% & 0.98 & 0.27 \\ \hline
$\Xi^{2S(Stein)}$(HRP) & 15.83\% & 18.23\% & 0.87 & -36.40\% & 0.97 & 0.26 \\ \hline
$\Xi^{2S(SymStein)}$(HRP) & 15.87\% & 18.16\% & 0.87 & -36.17\% & 0.98 & 0.27 \\ \hline
$\Xi^{2S(YCM)}$(HRP) & 15.69\% & 18.21\% & 0.86 & -36.21\% & 0.96 & $\mathbf{0.24}$ \\ \hline
\end{tabular}
\label{table5}
\end{table}

\section{Discussion}



We find that the estimator $\Xi^{[2S(YCM)]}$ substantially improves leverage and capital diversification, particularly in the moving window analysis for the minimum variance portfolio. This holds true both in the simulations of the complex and one-factor models, as well as in the empirical data when analyzing the MVP strategy at different sample sizes. Therefore, applying this two-step estimator can be beneficial in complex nested structures and those represented by factor models.  

We have also found that, for Model~1, it is difficult to achieve values of $\mathcal{RDI}$ less than 1. This indicates that in structures with a high degree of nested hierarchies, risk diversification becomes a challenging task, where even a uniform portfolio yields $\mathcal{RDI} = 1$. In contrast, for Model~3, it is possible to achieve much higher risk diversification, even with the classical high-dimensional linear estimator.


When comparing the estimators across different investment strategies, we observe that the artificial perturbations induced by the estimators successfully reduce the population metrics in some cases. Higher diversification is achieved in the complex model and the one-factor model for $\Xi^{2S(YCM)}$, regardless of the investment strategy. In general, the performance of $\mathcal{HHI}$ improves significantly with the HRP strategy for Models~1 and~2.  

Therefore, HRP is more suitable for complex structures, such as financial markets, in terms of improving capital diversification. In the case of leverage, the MVP+ and HRP strategies naturally yield the best results since, by construction, $\mathcal{L}=1$ regardless of the covariance matrix estimator. However, for the MVP strategy, $\Xi^{2S(YCM)}$ remains the best estimator.


Furthermore, in terms of $\mathcal{RDI}$, Model~1 under the MVP+ and HRP strategies does not benefit from the use of a sophisticated estimator; the sample estimator is sufficient.  
In fact, the $\mathcal{RDI}$ values for HRP are significantly lower than those for MVP and MVP+. This may be because the HRP strategy inherently aims to diversify risk.  
It can also be observed that $\mathcal{RDI}$ tends to decrease under the HRP strategy for Model~1. Therefore, for complex structures, HRP is the best investment strategy if the goal is to diversify risk.  
In Model~3, the best risk diversification is achieved by the two-step estimator optimized in terms of the Frobenius loss function, i.e., $\Xi^{2S(LP)}$.  
For its part, realized risk does not exhibit a clear pattern. However, it is worth noting that the lowest realized risk values are observed for the one-factor model under the MVP strategy, while the highest are seen for Model~1 under HRP.


Overall, the HRP strategy was found to be somewhat indifferent to the type of estimator used in the simulations. For this investment strategy, optimal values of risk diversification and realized risk were even achieved with the naive estimator in Model~1.  
Regarding the behavior of the metrics as a function of the sample size, it is observed that, in general, the naive estimator tends to converge to the values of the more sophisticated estimators as $n$ grows, while the latter exhibits a fairly moderate rate of decrease. Therefore, the sophisticated estimators remain optimal even for small $n$.  
Interestingly, it is observed that $\mathcal{HHI}$ increases as $n$ grows for Model~3, which is contrary to the behavior observed for all other metrics and analyzed cases.

For the empirical data, we found that its scree plot is similar to that of Model~1. From this, we infer a hierarchical and nested structure in the financial markets, as can be confirmed by examining its dendrogram.  
This finding is reinforced by observing a similar pattern in the capital diversification and leverage metrics. In the case of capital diversification, both Model~1 and the empirical data show the best performance with the estimator $\Xi^{2S(YCM)}$ for both MVP and MVP+. Interestingly, the magnitude of $\mathcal{HHI}$ in the empirical data is comparable to Model~1 under MVP and to Model~2 under MVP+, whereas the HRP values differ significantly.  
For leverage, again, Model~1 and the empirical data exhibit the best performance with the estimator $\Xi^{2S(YCM)}$, where only the MVP strategy is analyzed because it represents the non-trivial case. In this instance, the magnitude of the metric aligns more closely with Model~1.  
In terms of risk diversification, we observe that the empirical data's performance is closer to that of the one-factor model, as both have $\Xi^{YCM}$ as the best estimator. Here, the magnitude is also comparable to Model~2.  
For realized risk, it is observed that the empirical data, Model~1, and Model~2 achieve the best performance with non-linear estimators, at least for the MVP strategy. However, the magnitudes for the empirical data are significantly smaller.  
Overall, we observe stylized facts in the empirical data that lead us to hypothesize that the population structure of the covariance matrix is best characterized by a model lying between Model~1 and Model~2.


In the case of the walk-forward analysis, a notable observation is that the estimators do not significantly affect the behavior of the cumulative return under the HRP investment strategy. In fact, in all cases, HRP is the only strategy that closely approaches the performance of the uniform portfolio, which turns out to be the most profitable.  
This outcome is expected, as HRP is compared against minimum variance portfolios, and the estimators are designed to reduce noise in the risk estimation rather than maximize performance.


In fact, globally, after the uniform portfolio, the HRP strategy with the naive estimator is the combination that achieves the highest cumulative return in this walk-forward analysis for the empirical data.  
In contrast, the annual volatility reaches its lowest value with the estimator $\Xi^{2S(LP)}$ under the MVP+ strategy. On the other hand, the estimator $\Xi^{2S(YCM)}$ reduces the Maximum Drawdown for the MVP strategy. Turnover is also reduced with $\Xi^{2S(YCM)}$, but this time under the HRP strategy.  
In any case, applying a two-step estimator has a significant impact on the metrics, with notable implications for reducing risk and transaction costs. Meanwhile, the best balance between risk and performance was found to be associated with the estimator $\Xi^{YCM}$.


Interestingly, the HRP strategy is agnostic to most metrics with respect to the estimator considered. This allocation method demonstrated the greatest diversification both in the controlled experiments and in the empirical data.  

Moreover, risk diversification and realized risk were notably affected by the pandemic declaration event in March 2020. This date marks a turning point in the risk metrics, a behavior driven by heightened market volatility during that period of economic uncertainty.

\section{Conclusion}


The fully nested population covariance matrix models and the one-factor model capture stylized facts of financial markets from the perspective of portfolio theory.
The first model is proposed for the first time as a proxy for complex interactions, as it satisfies characteristics that define a complex system, such as the power law in its scree plot and nested hierarchies. The one-factor model, on the other hand, is supported by empirical evidence, appropriately explaining price formation, as interpreted by the Sharpe model\cite{molero2023market}.
Another key element of this study is the implementation of high-dimensional estimators to reduce noise in the estimation of the covariance matrix when the number of assets is comparable to the sample size.


Non-linear estimators of the covariance matrix were implemented, which are optimal for different loss functions. The estimator constructed through deterministic equivalents, as proposed by Yang et al.\cite{yang2015robust}, was also tested. One contribution in this direction is the implementation of its two-step version, a methodology originally proposed in \cite{garcia2023two}. In all cases, the estimators used contain mathematical elements with roots in different proof techniques within the theory of random matrices.
The performance of the estimators was also contrasted for different investment strategies.


In this paper, we conjecture that the fully nested hierarchical model adequately represents a complex system. However, it is necessary to study this model theoretically to understand its properties. The connection between the population eigenvalues of this system and the Fibonacci and Lucas numbers, as discussed in \cite{cahill2004fibonacci}, could provide guidance for a formal mathematical definition of a complex system in future work (see appendix~\ref{appendix}).


Our second conjecture is that financial markets exhibit stylized facts in portfolio theory that are shared with complex interaction models and one-factor models. These population models of the covariance matrix should be contrasted with different empirical datasets and under varying economic conditions to strengthen or refute this hypothesis in future work.


Finally, a third conjecture is that the two-step estimator $\Xi^{2S(YCM)}$ improves the diversification and leverage of complex and single-factor portfolios, both under controlled conditions and on empirical data. 
In its one-step version, the deterministic equivalents estimator and its properties are well-known from this proof technique based on random matrices. However, the theoretical properties of the two-step estimator are still unknown. Therefore, further study is needed to understand the effect of the second step (hierarchical estimators) on each of the one-step estimators, particularly on the versions of the non-linear estimators ($\Xi^{LP}$, $\Xi^{Stein}$, and $\Xi^{SS}$) and the context of the deterministic equivalents theory behind the $\Xi^{YCM}$ estimator.
Finally, it is important to integrate and study the two-step version of new proposals for random matrix estimators and machine learning methods, built from the promising and powerful technique of deterministic equivalents~\cite{bodnar2023dynamic,bodnar2024two}.

\section*{Acknowledgements}
This work was supported by National Council of Humanities, Sciences and Technologies (CONAHCYT, Mexico) [grant number CBF2023-2024-3976].
I thank Professor Rosario N. Mantegna for her invaluable support and motivation to develop this research project. We thank José Duarte Mendieta for providing the dataset from their thesis, which was instrumental in this study.

\appendix
\section{Characteristic polynomial of model 1}
\label{appendix}

The following population model is considered
\begin{equation}
    \mathbf{\Sigma} = \mathbf{LL^T}
\end{equation}
where the matrix $\mathbf{L}$ of dimension $p\times p$ is given by:\begin{equation}
    \mathbf{L} = \begin{pmatrix}
        \gamma & \gamma & \dots & \gamma & \gamma \\
        \gamma & \gamma & \dots &  \gamma & 0 \\
        \vdots &  \vdots & \ddots & \vdots & \vdots \\
        \gamma &  \gamma & \dots & 0 & 0 \\
        \gamma &  0 & \dots & 0 & 0 \\.
    \end{pmatrix}
\end{equation}
Hence
\begin{equation}
    \mathbf{\Sigma} = \mathbf{LL^T} = 
    \begin{pmatrix}
        p\gamma^2 & (p-1)\gamma^2 & (p-2)\gamma^2 & \dots & \gamma^2\\
        (p-1)\gamma^2 & (p-1)\gamma^2 & (p-2)\gamma^2 & \dots & \gamma^2\\
        (p-2)\gamma^2 & (p-2)\gamma^2 & (p-2)\gamma^2 & \dots & \gamma^2\\
        \vdots &    \vdots  & \vdots & \ddots & \vdots \\
        \gamma^2  & \gamma^2 & \gamma^2 & \gamma^2 &  \gamma^2
    \end{pmatrix}
\end{equation}

Specifically, the set of eigenvalues $\{\lambda\}$ of $\mathbf{\Sigma}$ can be found in linear time from the characteristic polynomial given by the determinant of the following tridiagonal symmetric Toeplitz matrix:

\begin{equation}
    det (\mathbf{\Sigma} - \lambda \mathbf{I}_p ) = 
    det \begin{pmatrix}
        \gamma^2 - 2\lambda & \lambda & 0 & \dots 0 & 0 \\
        \lambda  & \gamma^2 - 2\lambda & \lambda & \dots & 0\\
        0 & \lambda & \gamma^2 - 2\lambda & \dots & 0 \\
        \vdots & \vdots & \vdots & \ddots & \vdots \\
        0 & 0 & \lambda & \gamma^2 - 2\lambda & \lambda \\
        0 & 0 & 0 & \lambda & \gamma^2 - \lambda
    \end{pmatrix} = 0
\end{equation}

\bibliography{references}
\end{document}